# Design and evaluation of conjugated polymers with polar side chains as electrode materials for electrochemical energy storage in aqueous electrolytes

**Authors:** Davide Moia,*,a,‡ Alexander Giovannitti,*,a,b,‡ Anna A. Szumska,a Iuliana P. Maria,b Elham Rezasoltani,a Michael Sachs,b Martin Schnurr,b Piers R.F. Barnes,a Iain McCulloch,b,c Jenny Nelson*,a

**Affiliations**

a Department of Physics, Imperial College London SW7 2AZ London, UK

b Department of Chemistry, Imperial College London SW7 2AZ London, UK

c Physical Sciences and Engineering Division, KAUST Solar Center (KSC), King Abdullah University of Science and Technology (KAUST), KSC Thuwal 23955-6900, Saudi Arabia

* davide.moia11@imperial.ac.uk; a.giovannitti13@imperial.ac.uk; jenny.nelson@imperial.ac.uk

‡ These authors contributed equally to this work

**Keywords:** redox-active polymers, post-functionalization reactions, single-phase polymer electrodes, fast charging, safe electrolytes.

**Abstract**

We report the development of redox-active conjugated polymers with potential application to electrochemical energy storage. Side chain engineering enables processing of the polymer electrodes from solution, stability in aqueous electrolytes and efficient transport of ionic and electronic charge carriers. We synthesized a 3,3'-dialkoxybithiophene homo-polymer (p-type polymer) with glycol side chains and prepared naphthalene-1,4,5,8-tetracarboxylic-diimide-dialkoxybithiophene (NDI-gT2) copolymers (n-type polymer) with either a glycol or zwitterionic side chain on the NDI unit. For the latter, we developed a post-functionalization synthesis to attach the polar zwitterion side chains to the polymer backbone to avoid challenges of purifying polar intermediates. We demonstrate fast and reversible charging of solution processed electrodes for both the p- and n-type polymers in aqueous electrolytes, without using additives or porous scaffolds and for films up to micrometers thick. We apply spectroelectrochemistry as an *in operando* technique to probe the state of charge of the electrodes. This reveals that thin films of the p-type polymer and zwitterion n-type polymer can be charged reversibly with up to two electronic charges per repeat unit (bipolaron formation). We combine thin films of these polymers in a two-electrode cell and demonstrate output voltages of up to 1.4 V with high redox-stability. Our findings demonstrate the potential of functionalizing conjugated polymers with appropriate polar side chains to improve specific capacity, reversibility and rate capabilities of polymer electrodes in aqueous electrolytes.

**Introduction**

Polar side chains attached to conjugated polymer backbones have been investigated as a strategy to facilitate ionic transport in the bulk of mixed electronic-ionic conducting polymers.[1–4] In particular, ethylene glycol based side chains have been used to facilitate ion transport in conjugated polymer films leading to the successful demonstration of organic electrochemical transistors (OECTs),[3–5] electrochromic[6,7] and biomedical devices[8] operating in aqueous electrolytes. Conjugated polymers with polar side chains are therefore a promising class of materials for the development of electrodes for safe and sustainable electrochemical energy storage devices using water based electrolytes.

Prior attempts to develop polymer based electrodes with high energy and power densities have typically targeted polymers with a high density of redox sites in combination with conductive additives or scaffolds and inert polymer binders.[9,10,11,12] Most studies report operation of these electrodes in organic electrolytes that show a wide electrochemical stability window but are usually hazardous and/or flammable. Several examples of conjugated p-type polymers have been previously presented, showing in some cases competitive specific capacities. Significantly fewer reports of n-type conjugated polymer electrodes have been made and, to our knowledge, no previous study successfully demonstrated this class of materials as electrodes in water-based applications. Liang *et al*. have shown that naphthalene-1,4,5,8-tetracarboxylic-diimide-bithiophene (NDI-T2) copolymers with alkyl side chains can be a promising electrode material (capacity of ~50 mAh/g and charging times in the order of ~10 seconds) when the copolymer is mixed with conductive carbon and measured in an organic electrolyte.[13] Functionalizing conjugated polymer backbones with alkyl side chains increases their solubility in common organic solvents and facilitates processing of electrodes using printing or coating techniques. However, alkyl chains are an unsuitable transport medium for polar hydrated ions[3] as shown by previous studies on poly(3-hexylthiophene-2,5-diyl) (P3HT)[14] and naphthalene1,4,5,8-tetracarboxylic-diimide-bithiophene copolymers (NDI-T2).[4]

To achieve the levels of ionic conductivity, specific capacity and stability that are needed for the application of conjugated polymers as electrodes, the polymer backbone and side chains need to be optimized independently. Redox-stability in aqueous electrolytes requires that the oxidation and reduction potentials of the electrode materials lie at voltages within the electrochemical stability window of water.[15] Electron-rich polythiophene (p-type), and electron-deficient donor-acceptor and acceptor-type polymer backbones (n-type) show, respectively, low oxidation potentials and low reduction potentials in aqueous electrolytes which makes them suitable structures for electrochemical charging in water. Efficient ionic exchange with aqueous electrolytes has been achieved for p-type polymers based on polythiophenes by using glycol side chains.[3] On the other hand, the use of glycol chains in n-type polymers based on NDI-T2 results in low electron mobility, as shown by measurements on OECT devices, and is also expected to inhibit cation transport.[16] Alternative polar side chains, such as zwitterionic and polyelectrolyte, have been attached to NDI-T2 copolymers to improve interfacial properties in organic electronic devices but have not so far not been tested in electrochemical energy storage devices.[17,18] Polyelectrolytes with a mobile cation have been explored as side groups of non-conjugated radical polymers, where the cation is exchanged with the electrolyte to form a zwitterion structure when the polymer is oxidized. This strategy improved the reversibility as well as the capacity of the polymer in water electrolytes, suggesting that the interaction between the polymer redox sites and compensating ionic charge plays an important role on the electrochemical response of the material.[19] Appropriate choice of the side chain is expected to also influence the reversibility and specific capacity of n-type polymer structures, but design rules that address this question are currently missing. In addition, while polar side chains are known to facilitate ion transport in polymer films, it is not clear to what extent this strategy can improve the rate capability of compact, solution processed polymer electrodes of micrometer scale thickness.

Here, we report the development of solution processable redox-active polymers with the goal of enabling electrochemical energy storage in aqueous electrolytes. We choose polymer backbones which show high stability during electrochemical redox-reactions and engineer the side chains to enable reversible charging in water based electrolytes. We synthesize a p-type polymer based on a homo-3,3'-dialkoxybithiophene polymer (p(gT2), Figure 1a) with methyl end-capped triethylene glycol side chains (TEG). For the n-type polymer, we prepare donor-acceptor type copolymers with naphthalene-1,4,5,8-tetracarboxylic diimide (NDI) and 3,3'-dialkoxybithiophene[20–22] where we either attach zwitterionic side chains (p(ZI-NDI-gT2), Figure 1b) or linear glycol side chains (p(g7NDI-gT2)[4] to the NDI unit. For the zwitterionic side chain, we place the ammonium ion in close proximity (C2-spacer) to the NDI unit with the aim of enabling the electronic charge to interact with the ammonium ion during charging of the polymer rather than requiring charge compensating cations to approach the polymer backbone (Figure 1c). For all n-type copolymers reported here, TEG side chains are attached on the donor comonomer to ensure high solubility of the polymers in common organic solvents, which allows solution processing of the materials. We characterize the polymer films in sodium chloride aqueous solution using electrochemistry and spectroelectrochemistry, revealing their ability to host bipolarons and to charge/discharge on the second timescale for > 1 micron thick samples. We finally explore their application in a two electrode electrochemical energy storage device, which can be charged up to 1.4 V, and discuss some of the limitations involved when considering salt water as the electrolyte of the cell.

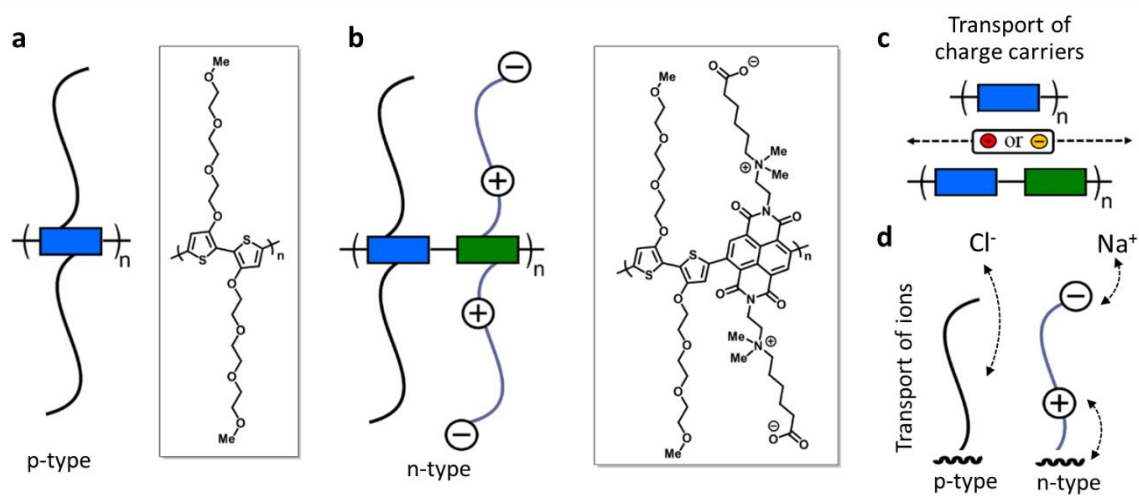

Figure 1. a) p-type homo-polymer with TEG side chains p(gT2); b) n-type donor-acceptor copolymer with zwitterion side chains on the acceptor units (marked as green blocks in the schematic) and TEG side chains on the donor units (blue blocks) p(ZI-NDI-gT2). Schematic of the transport of c) the electronic charges along the backbone of the polymers and d) ions with the aid of polar side chains.

**Results and discussion**

The properties of the p- and n-type polymers are summarized in Table 1. Details on synthesis and characterization of monomers and polymers are reported in Section 1-7 of the ESI. Polymer p(gT2) was synthesized by Stille polymerization as previously reported for other copolymers containing glycol side chains.[23] The polymer thin film absorption spectrum is presented in Figure 2a and shows an absorption onset of 755 nm with two vibronic transitions (0-0 and 0-1). The ionization potential (IP) of p(gT2) was measured by photoelectron spectroscopy in air (PESA, 4.5 eV) and cyclic voltammetry (CV, 4.5 eV) in organic electrolytes where the IP of the polymer was calculated as reported in the literature[20,24] (Figure 2b). The molecular weight distribution of the polymer was measured by gel permeation chromatography (GPC) in DMF ($M_n$ = 36 kDa and $M_w$ = 60 kDa). To avoid overestimation of the molecular weight distribution mainly due to aggregation of the conjugated polymer with glycol side chains,[4] the high molecular weight fraction of the GPC trace was neglected and only the lower molecular weight fraction was considered for molecular weight analysis (Figure S10, bimoduale eluation). Additionally, mass spectrometry measurements by matrix-assisted laser deposition/ionization time of flight (MALDI-ToF) were carried out and polymer chains length with >35 repeat units were detected (Figure S11, (corresponding to >18 kDa)). In comparison to other reported 3,3'-dialkoxybithiophene copolymers with glycol side chains on only every other repeat unit[3,23], p(gT2) has a higher solubility in organic solvents such as N,N-dimethylformamide (DMF) or chloroform ($CHCl_3$). Good solubility of the polymer is essential for fabrication of electrodes from solution.

The n-type polymer p(ZI-NDI-gT2) was synthesized following a post-functionalization route as shown in Figure 3a where the precursor copolymer p((DMA)-NDI-gT2) with dimethylamino groups was prepared by Stille polymerization. The dimethylamino groups of p((DMA)-NDI-gT2) were first converted into ammonium bromides featuring an ester group at the end of the side chain. $^1$H Nuclear magnetic resonance (NMR) spectroscopy and 2D correlation spectroscopy (COSY) were carried out to monitor the formation of the ammonium bromide (Figure S16). In the next step, the ester was cleaved to form the ammonium-carboxylate zwitterionic copolymer p(ZI NDI-gT2). Finally, the polymer was purified by dialysis in deionized water to remove water soluble side products. $^1$H NMR spectra (Figure 3b) were recorded to monitor the ester cleavage, showing the disappearance of the signals corresponding to the proton of the ethyl group of the ester (4.1 ppm and 1.2 ppm) in accordance with reactions reported in the literature.[25] This shows that the here presented post-functionalization approach can be a viable route to attach functional groups at NDI copolymers. The UV-Vis spectrum of p(ZI-NDI-gT2) is shown in Figure 2a where the copolymer has two absorption peaks and an absorption onset of 1049 nm. The IP of p(ZI-NDI-gT2) was measured to be 5.15 eV (5.20 eV by CV, Figure 2b). The molecular weight distribution of p(ZI-NDI-gT2) was measured by GPC in DMF ($M_n$ = 24 kDa and $M_w$ = 53 kDa), which most likely overestimates the molecular weight distribution of the polymer since only oligomers up to 6 repeat units could be detected for the precursor copolymer p((DMA)-NDI-gT2) (see Figure S13).

Unfortunately, MALDI-ToF measurements of p(ZI-NDI-gT2) were inconclusive. Polymer p(ZI-NDI-gT2) is soluble in polar organic solvents such as dimethyl sulfoxide (DMSO) and methanol (MeOH). Note that the substitution of the anion from carboxylate to sulfonate resulted in the formation of a water soluble polymer which could therefore not be tested as an electrode material in aqueous electrolytes. The n-type polymer heptakis-ethylene glycol side chain on the NDI unit (p(g7-NDI-gT2)) was synthesized by Stille coupling and the properties of the polymers are reported in section 5 of the ESI.

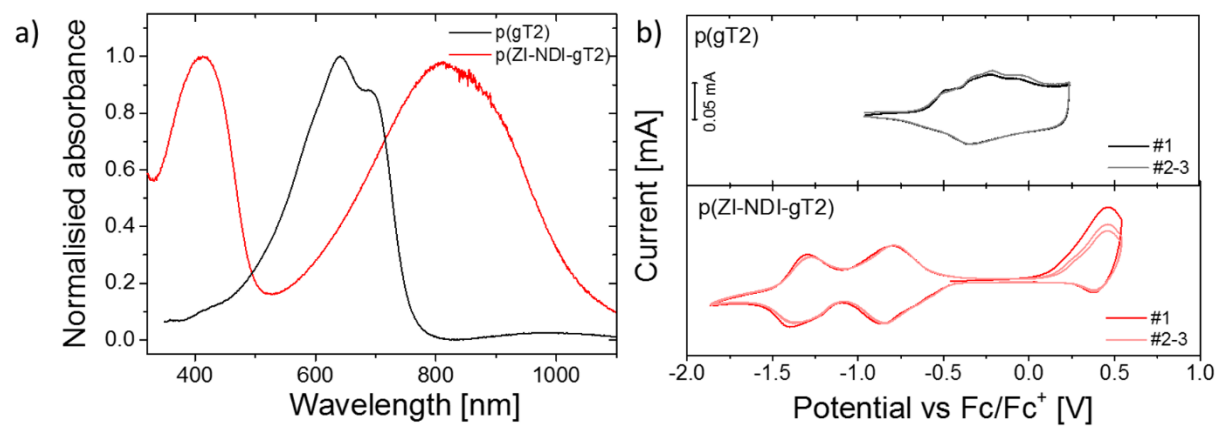

Figure 2. a) Thin film UV-Vis spectra of a p(gT2) film on fluorine doped tin oxide (FTO) glass and of a p(ZI-NDI-gT2) film on a glass substrate. A potential of -0.3 V vs Ag/AgCl was applied prior to recording of the UV-Vis spectrum (three-electrode setup) (p(gT2) can become oxidized in ambient conditions). b) Thin film CV measurements of p(gT2) and p(ZI-NDI-gT2) on FTO substrates were recorded by using a degassed 0.1 M NBu$_4$PF$_6$ acetonitrile solution as supporting electrolyte (100 mV/s, three cycles (#1-3) are presented where the first cycle (#1) is highlighted).

Table 1. Summary of the polymers' properties

| Polymer | IP[A] [eV] | IP[B] [eV] | Absorption onset [nm] | Optical band gap [eV] | Mn[D] [kDa] | Mw[D] [kDa] |
|---|---|---|---|---|---|---|
| p(gT2) | 4.48 | 4.50 | 755 | 1.64 | 36 [E] | 60 [E] |
| p(ZI-NDI-gT2) | 5.15 | 5.25 | 1049 | 1.18 | 24 | 53 |

[A] Photoelectron spectroscopy in Air (PESA)

[B] CV measurements were carried out in acetonitrile (0.1 M NBu$_4$PF$_6$, 100 mV/s)

[C] GPC measurements were carried out in DMF with 5 mM NH$_4$BF$_4$.

[D] Bimoduale eluation, high molecular weight fraction (aggregates) were not considered in the molecular weight analysis.

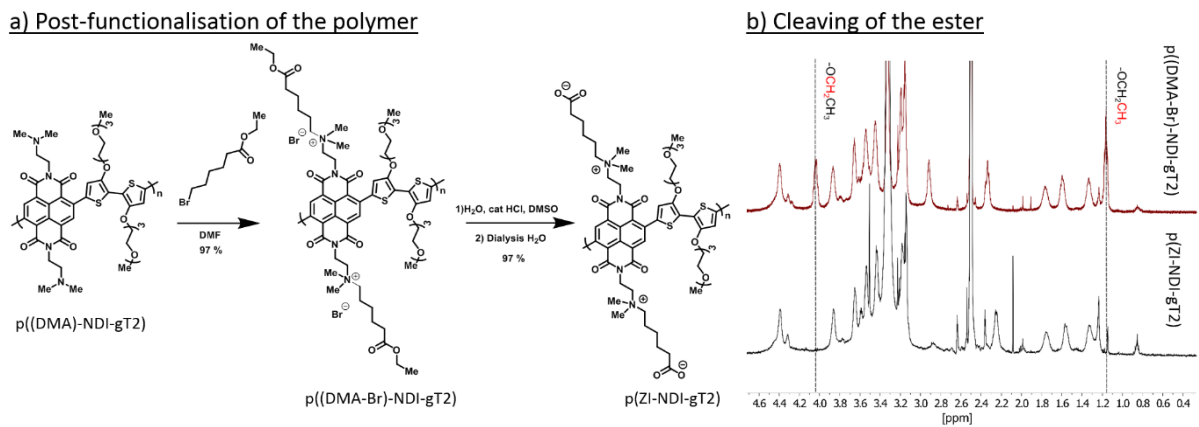

Figure 3. a) Post-functionalization of p((DMA)-NDI-gT2) to form p(ZI-NDI-gT2) and b) $^1$H NMR spectra of p((DMA-Br)-NDI-gT2) and p(ZI-NDI-gT2), highlighting (vertical lines at 4.04 ppm and 1.16 ppm) the protons corresponding to the ester of p((DMA-Br)-NDI-gT2) and the disappearance after cleavage.

We now consider the electrochemical and spectroelectrochemical characterization of thin films of the polymers described above in aqueous electrolytes. Figure 4 a and b show cyclic voltammetry measurements of thin films of both p-type (p(gT2)) and n-type (p(ZI-NDI-gT2)) polymers performed in 0.1 M NaCl aqueous solution in a three electrode cell (more details on the experimental conditions are described in Section 8 of the ESI). Similarly to the charging in organic electrolytes, both polymers show highly reversible redox features with stable peak positions after the first cycle. The polymer p(gT2) shows two reversible oxidation peaks at relatively low potentials, consistent with previous reports on electrochemistry of polythiophene based polymer films.[26] Remarkably, the reduction of p(ZI-NDI-gT2) in the aqueous electrolyte also shows two distinct reduction peaks. This suggests that the NDI unit can be reversibly charged with two electrons in water, in agreement with findings in the literature.[13,27] Continuous cycling of polymer films at 25 mV s$^{-1}$ showed that for p(gT2) and p(ZI-NDI-gT2), 98% and 63 % of the initial (2$^{nd}$ scan) capacity is retained after 500 cycles in 0.1 M NaCl aqueous solution (Figure S37).

Since conjugated polymers show a distinct color change between their neutral and charged state, we used spectroelectrochemical measurements to monitor the optical absorbance of the films during the CV measurements shown in Figure 4a and b. Figure 4c shows that the main absorption feature of p(gT2) in the visible region (peak at 645 nm) is completely quenched upon oxidation of the polymer by applying a potential of up to 0.2 V vs Ag/AgCl and that a new absorption band appears in the NIR. In agreement with reported results for other conjugated polymers,[28] we attribute the change of the polymer neutral state absorption to the formation of positive polarons on the polymer backbone. Further oxidation ($V > 0.2$ V vs Ag/AgCl) reduces the NIR absorption suggesting that further conversion occurs, which we ascribe to bipolaron formation.[29] Calculated spectra of the trimer (gT2)$_3$ in its neutral, singly and doubly charged state using time-dependent density functional theory (TD-DFT) shown in Figure 4e (further details are presented in Section 9 of the ESI) support our assignment of the spectral changes to polaron and bipolaron formation, as do prior reports on polythiophene spectroelectrochemistry.[23,29,30] Importantly, the spectral evolution suggests complete conversion of neutral polymers into the charged state, indicating that the full volume of the electrode undergoes charging. Scanning to more positive potentials shows that higher levels of charging can be achieved, although this compromises the coulombic efficiency and stability of the electrode (see section 10 of the ESI). The results for the n-type polymer p(ZI-NDI-gT2) are presented in Figure 4d and indicate reversible formation of an electron polaron at a potential of -0.4 V and bipolaron between -0.4 V and -0.75 V vs Ag/AgCl. Our assignment of the spectral features to polaron and bipolaron formation are again supported by calculated absorbance spectra of neutral and charged monomers (Figure 4f). The capacity of the redox-active polymers in aqueous electrolytes was measured to be 25 mAh cm$^{-3}$ and 36 mAh cm$^{-3}$ for p(gT2) and p(ZI-NDI-gT2), respectively (see section 10 and 11 of the ESI). Measurements of gravimetric capacities are also shown in Figure S38. We note that for these estimates we are using the volume (mass) of the dry films and neglect volume expansions (mass increase) through swelling in the presence of the aqueous electrolyte.

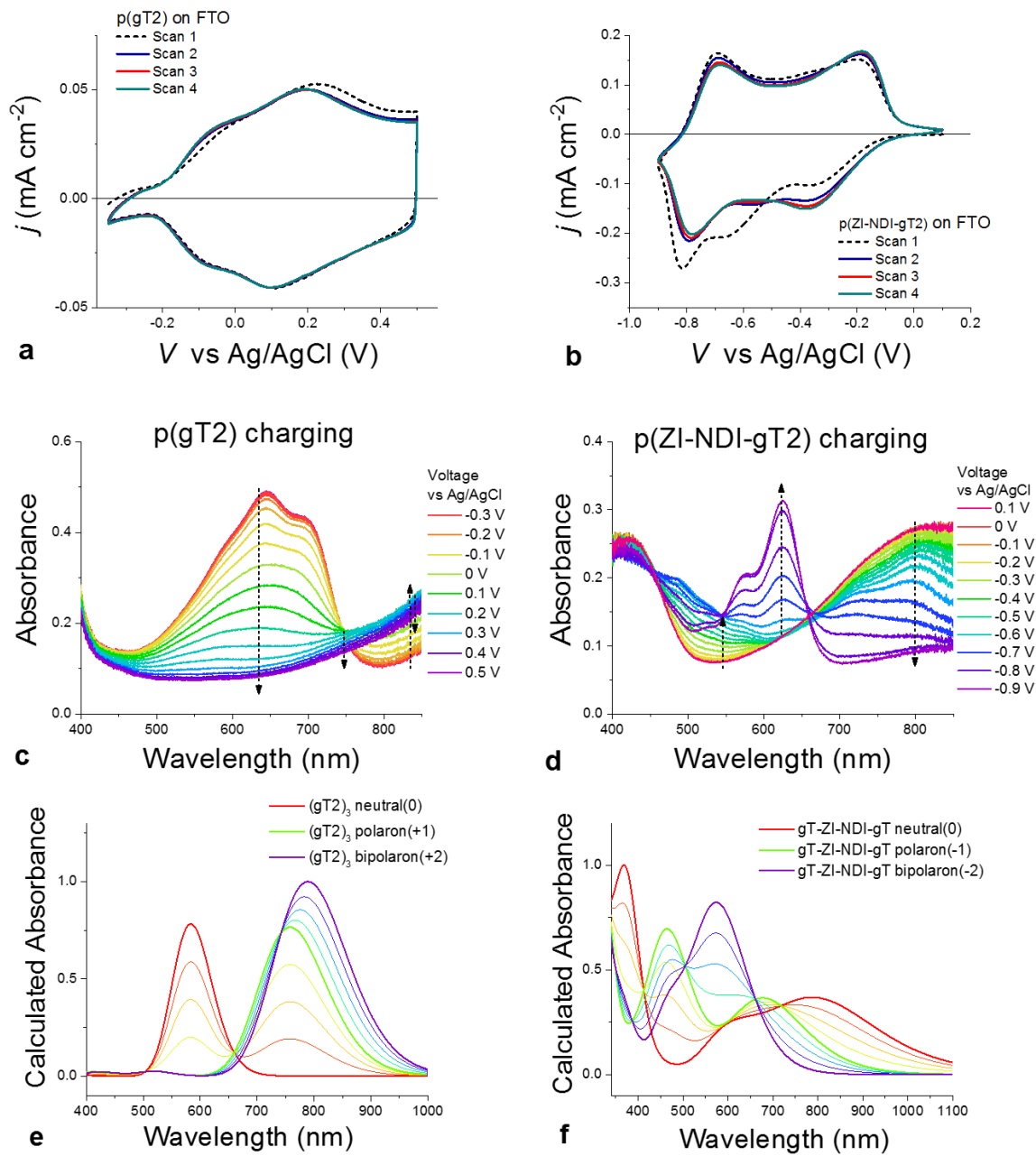

Figure 4. Characterization of p(gT2) (left) and p(ZI-NDI-gT2) (right). CV measurements of a) p(gT2) and b) p(ZI-NDI-gT2) using scan rate of 50 mV s$^{-1}$ (cycle #1-4) . To avoid side reactions with oxygen, the electrolyte for the n-type polymer was degassed with argon for 15 min prior to performing the measurement. UV vis absorbance spectra of c) p(gT2) and d) p(ZI-NDI-gT2) for the charging of the polymer films during the first CV scan shown in (a, b). Normalised absorbance spectra, calculated with TD-DFT for e) the trimer (gT2)$_3$ and f) the monomer gT-ZI-NDI-gT in the neutral and charged (polaron and bipolaron) states (bold lines). Linear combinations of these three spectra were used to plot spectra of intermediate states (light lines).

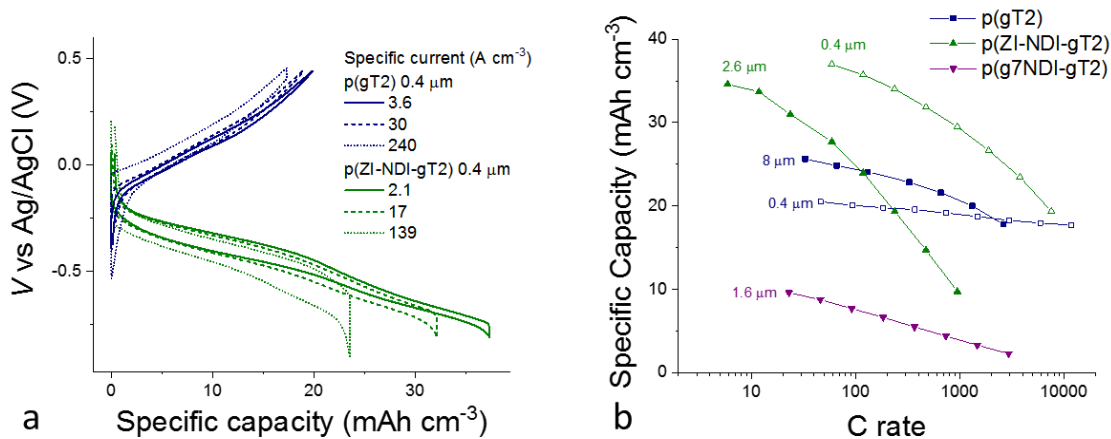

Figure 5. a) Galvanostatic charge–discharge profiles of p(gT2) (blue curves) and p(ZI-NDI-gT2) (green curves) at different specific currents and b) Specific capacity as a function of C rate of the p-type and n-type polymers on gold coated glass substrates with a degassed 5 M NaCl aqueous solution as the supporting electrolyte. For both graphs, the datasets correspond to the second measurement performed at each specific current (C-rate) condition. The data in (b) display the charge obtained during the discharging of the electrode. Additional data including measurements on the films at lower electrolyte concentration are reported in Figure S40 – S44.

The rate capabilities of the electrode materials were studied in NaCl aqueous electrolytes by performing galvanostatic charge/discharge measurements of films of between 400 nm and several microns thickness. The results are summarized in Figure 5. For these measurements, we used gold coated glass substrates and a 5 M NaCl aqueous solution to minimize the series resistance of the cell. The galvanostatic charge–discharge profiles during charging of 400 nm thick p(gT2) and p(ZI-NDI-gT2) films are shown in Figure 5a. Polymer p(gT2) can be charged reversibly to 0.5 V vs Ag/AgCl and its specific capacity drops by less than 15 % when the current density is increased from 3.6 A/cm³ to 240 A/cm³. For p(ZI-NDI-gT2), the specific capacity drops by 37 % when the current density is increased from 2.1 A/cm³ to 139 A/cm, suggesting lower rate capabilities than the p-type polymer considered here. The p(ZI-NDI-gT2) polymer does not show further charging at more negative voltages, after the formation of the bipolaron at -0.85 V vs Ag/AgCl, which is in agreement with previous reports on NDI-T2 copolymers.[13]

Figure 5b presents the rate capabilities of thick electrode films of p(gT2) and p(ZI-NDI-gT2) as well as showing the influence on rate capability of replacing the zwitterionic side chains on p(ZI-NDI-gT2) with glycol chains (p(g7NDI-gT2) polymer. For p(gT2), we observe that the specific capacity for a 8 µm thick film drops by less than 20% when increasing the C rate from 30 to 1000 C (specific capacity per unit area of 0.02 mAh cm⁻²). This remarkable finding shows that, for p(gT2), ionic transport occurs in the second timescale even for thicknesses on the order of 10 µm. Similarly fast kinetics for solution processed polymer films in aqueous electrolytes were previously reported only for thin films (~100 nm)[31] or microns thick p-type polymer films processed through additional acid/base treatments.[32,33] The n-type polymer p(ZI-NDI-gT2) shows higher specific capacity than p(gT2) at low charging rates, however it also shows lower rate capabilities as shown in Figure 5a. Interestingly, the n-type polymer with zwitterionic side chains on the NDI repeat unit (p(ZI-NDI-gT2)) shows a more than three times higher specific capacity than the same n-type backbone with linear glycol side chains (p(g7NDI-gT2)). For polymer p(g7NDI-gT2), we observe an upper limit to reversible charging when scanning to potentials beyond the first reduction peak (Figure S36 and S37). We hypothesize that this observation is due to stronger interaction of the glycol side chains on the NDI unit with alkali-metal ions[16] and water molecules[4] compared to the ZI side chain. Such interactions might also induce a more pronounced uptake of water in the film and induce the observed faster capacity fading of thin films of this polymer during continuous cycling. In particular, during charging of the polymer, additional water molecules migrate into the structure as part of the alkali metal ions' hydration shell where water molecules can then also interact with the glycol side chains via hydrogen bonding. As a result, more water molecules remain in the polymer structure after discharging of the polymer, and could underlie the observed change in electrochemical response of the film.

We note that the specific capacity per unit area of the thickest film reported here (0.02 mAh cm$^{-2}$) is low compared to other reported electrochemical energy storage electrodes using non-conjugated organic materials with no side chains processed with conductive additives.[34] This is a direct consequence of using redox-inactive polar side chains to enable fast charging of the electrodes in neutral aqueous electrolytes. However, the observation that side chain engineering of the n-type polymer can be used to improve the electrode specific capacity suggests that side chain optimization could allow to reach the theoretical specific capacity of these materials even in non-porous single-phase films, beside enabling solution processing of the polymers.

Based on the promising results obtained from the galvanostatic measurements for individual electrodes, thin films of p(ZI-NDI-gT2) and p(gT2) with similar capacity were prepared and tested in a two-electrode electrochemical cell with a 0.1 M NaCl aqueous solution as the supporting electrolyte. The charging process of the two-electrode cell is illustrated in Figure 6a. When a positive potential $V_{cell} > 0$ V is applied to the cathode with respect to the anode, Na+ ions (Cl- ions) from the electrolyte migrate into the bulk of the polymer to compensate for the negative (positive) electronic charge on the p-type (n-type) polymer backbone. The redox reactions of both polymers with corresponding potentials vs Ag/AgCl are expressed in Figure 6b, where we also illustrate the electrochemical window of neutral water in ambient conditions including electrochemical redox-reactions involving oxygen. Electrochemical side reactions, including reduction of oxygen and water splitting, can occur at relatively low potentials in aqueous electrolytes and it is therefore important to design the electrode materials accordingly to avoid them.

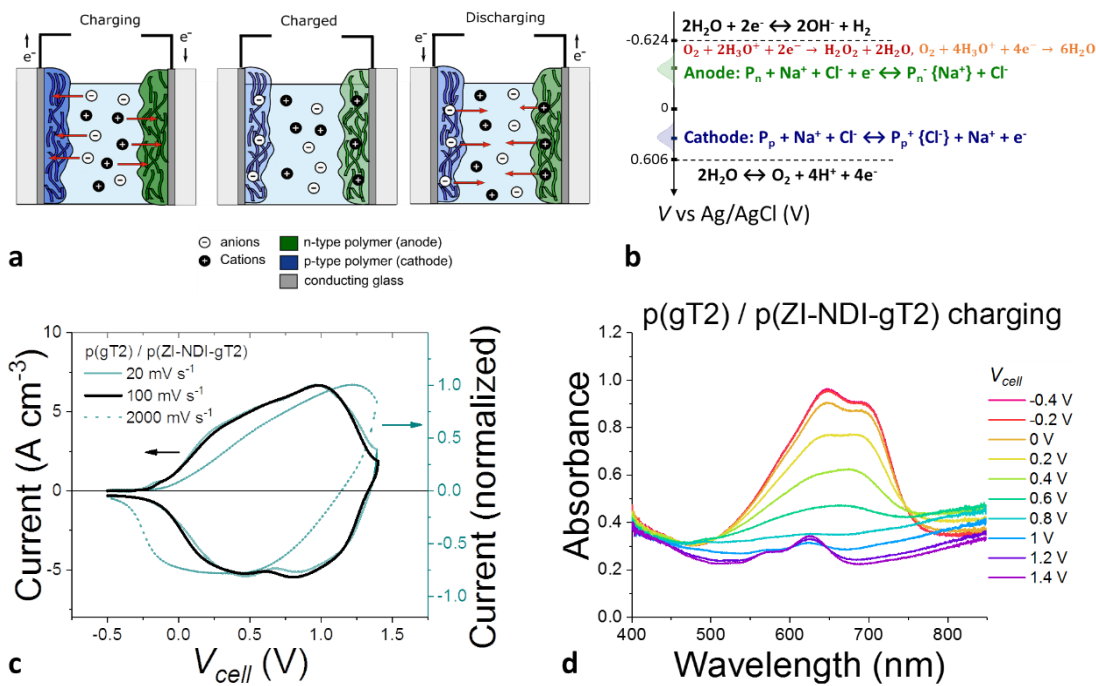

Figure 6. (a) Schematics of structure and charge distribution at different charge states of the electrochemical energy storage cell using films of the p- and the n-type polymers as the two electrodes, separated by an aqueous electrolyte. (b) Reactions at the cathode and at the anode and electrochemical window of water at neutral pH. Characterization of a two electrode cell with structure FTO / p(gT2) (87 nm)/ 0.1 M NaCl:DIW / p(ZI-NDI-gT2) (70 nm) / FTO. The voltage ($V_{cell}$) is applied/measured at the cathode (p(gT2) electrode) with respect to the anode (p(ZI-NDI-gT2) electrode). (c) Cyclic voltammetry measurements performed at different scan rates. (d) Evolution of the optical absorbance for the two films when charging the cell to 1.4 V at 100 mV s$^{-1}$.

The CV measurements of the cell are presented in Figure 6c. The device is charged up to 1.4 V, which corresponds to the potential range used to characterize the p-type (0.5 V vs Ag/AgCl) and n-type (-0.9 V vs Ag/AgCl) polymers (see also Figure S45). The spectroelectrochemical response resembles the contribution of the individual electrode spectra and shows that electrochemical charging of the cell produces the bipolaron in both films as shown separately for the two films (Figure 4c, d). Further characterization of the cell is shown in section 15 of the ESI. We carried out charge-retention experiments in degassed 0.1 M NaCl aqueous solutions to investigate electrochemical side-reactions of the charged electrodes and the aqueous electrolyte. During this experiment, we observed a charge retention of 25% after 5000 seconds (see Figure S47 and S48), and discovered that oxygen in the aqueous electrolyte plays an important role, influencing the retention of electronic charges in the electrodes. When oxygen is present in the electrolyte, a larger charge is measured during charging than during discharging of the cell, which results in a lower coulombic efficiency and can compromise the stability of the polymer films. Electron transfer from either the neutral p-type p(gT2) or the reduced n-type polymer p(ZI-NDI-gT2) to oxygen is an expected side reaction,[15] where oxygen can either be converted into hydrogen peroxide (two electron process) or water (four electron process).[15,35,36] Based on recent findings of efficient electrocatalytic production of hydrogen peroxide of electrodes made of organic polymers,[37] we hypothesized that one pathway for the loss of the charges is an electron transfer from the reduced n-type polymers to oxygen dissolved in the electrolyte. To test this hypothesis, we charged the electrochemical cell in ambient conditions and used the horseradish peroxidase/3,3′,5,5′-tetramethylbenzidine (dye/enzyme) system[38] to detect potential generated hydrogen peroxide. We observed hydrogen peroxide formation after charging the electrode in the presence of oxygen and the results are presented in Figure S50. Additionally, we monitored hydrogen and oxygen evolution in degassed aqueous electrolyte to verify whether other faradaic electrochemical side reactions such as water splitting occur in our system and observed no changes of oxygen or hydrogen concentrations during the charging of the cell (Figure S49). In order to study the origin of the loss of charge in the electrochemical cell in more detail, we use spectroelectrochemistry to monitor the changes of the absorption spectrum during the retention experiment. We observed the disappearance of the absorption features of the bipolaron on the n-type polymer when the cell is held at open circuit in the charged state. Based on this and the observation described above, we assign the charge retention limitation of the cell to the formation of hydrogen peroxide from the reaction of electrons on the n-type polymer and residues of oxygen left in the electrolyte (see reactions in Figure 6b and data in Figure S50). As a result, we expect the p-type polymer to accumulate holes upon cycling of the cell, as confirmed by spectroelectrochemical measurements and by the drop in capacity illustrated in Figure S51. This drop can be partially recovered by applying a negative bias (Figure S52). We finally note that the observed charge retention time is comparable to measurements carried out for other electrodes based on organic semiconductors in organic electrolytes and shows that oxygen free aqueous electrolytes can be an interesting electrolyte for the development of electrochemical energy storage devices.[10]

**Conclusion**

We demonstrated the development of solution processable, fast switchable p- and n-type electrode materials to operate in water, offering a route towards electrochemical energy storage in safe and sustainable electrolytes. Ethylene glycol based side chains attached to a p-type polymer backbone enabled reversible redox-reactions in aqueous electrolytes and fast charging of the polymer films in the second timescale for up to 8 µm thick films. For NDI-T2 n-type polymers we observed limited reversibility when using glycol side chains. We presented a chemical design strategy to prepare an n-type polymer with zwitterionic side chains via a post-functionalization reaction route and demonstrated improvement in both specific capacity and reversibility of the polymer electrode in aqueous electrolytes. When combined in a two-electrode cell, our p-type and n-type polymers showed reversible charging up to 1.4 V in a neutral aqueous sodium chloride salt solution. We used spectroelectrochemistry as an *in-operando* measurement tool to monitor electrochemical charging/discharging of semitransparent electrodes and observed that both polymers formed reversible bipolaron states, approaching their theoretical capacities. Additionally, we used the technique to monitor the state-of charge of the cell as well as to highlight the charge retention limitations of the n-type polymer. The study illustrates the potential and some of the limitations of using aqueous electrolytes as a safe and sustainable solution for energy storage applications. The water switchable redox-materials reported here will also be applicable to novel devices for bio-sensing, electrochromic and memory applications, where control of mixed ionic-electronic conduction plays an important role.

**Conflicts of interest**

There are no conflicts to declare.

**Acknowledgements**

We thank Peter R Haycock for the fruitful discussions about the NMR spectra. Funding: DM, PB and JN are grateful for funding from the EPSRC Supersolar Hub (grant EP/P02484X/1). This project has received funding from the European Research Council (ERC) under the European Union's Horizon 2020 research and innovation program (grant agreement No 742708) and EC H2020 Project SOLEDLIGHT (grant agreement No 643791), IMEC Synergy Grant SC2 (610115), EPSRC Project EP/G037515/1, EP/M005143/1, EP/N509486/1 and from The Imperial College Faculty of Natural Sciences Strategic Research Fund.

**Supplementary information**

**Design and evaluation of conjugated polymers with polar side chains as electrode materials for electrochemical energy storage in aqueous electrolytes**

**Authors:**

**Authors:** Davide Moia,*,[a,‡] Alexander Giovannitti,*,[a,b,‡] Anna A. Szumska,[a] Iuliana P. Maria,[b] Elham Rezasoltani,[a] Michael Sachs,[b] Martin Schnurr,[b] Piers R.F. Barnes,[a] Iain McCulloch,[b,c] Jenny Nelson*,[a]

**Affiliations**

[a] Department of Physics, Imperial College London SW7 2AZ London, UK

[b] Department of Chemistry, Imperial College London SW7 2AZ London, UK

[c] Physical Sciences and Engineering Division, KAUST Solar Center (KSC), King Abdullah University of Science and Technology (KAUST), KSC Thuwal 23955-6900, Saudi Arabia

* davide.moia11@imperial.ac.uk; a.giovannitti13@imperial.ac.uk; jenny.nelson@imperial.ac.uk

[‡] These authors contributed equally to this work

# 1. Table of Contents

# 2. Material synthesis and characterization

Column chromatography with silica gel from VWR Scientific was used for flash chromatography. Microwave experiments were carried out in a Biotage Initiator V 2.3. $^1$H and $^{13}$C NMR spectra were recorded on a Bruker AV-400 spectrometer at 298 K and are reported in ppm relative to TMS. UV-Vis absorption spectra were recorded on UV-1601 ($\lambda_{max}$ 1100 nm) UV-VIS Shimadzu spectrometers.

MALDI TOF spectrometry was carried out in positive reflection mode on a Micromass MALDImxTOF with trans-2-[3-(4-tert-Butylphenyl)-2-methyl-2-propenylidene]-malononitrile (DCTB) as the matrix.

Cyclic voltammograms were recorded on an Autolab PGSTAT101 with a standard three-electrode setup with ITO coated glass slides, a Pt counter electrode and a Ag/AgCl reference electrode (calibrated against ferrocene (Fc/Fc$^+$)). The measurements were either carried in an anhydrous, degassed 0.1 M tetrabutylammonium hexafluorophosphate (TBAPF$_6$) acetonitrile solution or in a 0.1 M NaCl aqueous solution as the supporting electrolyte with a scan rate of 100 mV/s. Ionisation potential were calculated according to the literature: IP [eV] = $E_{(Ox,onset\ vsFc/Fc+)}$+5.1.[1]

Gel permeation chromatography (GPC) measurements were performed on an Agilent 1260 infinity system operating in DMF with 5 mM NH$_4$BF$_4$ with 2 PLgel 5 µm mixed-C columns (300×7.5mm), a PLgel 5 mm guard column (50x7.5mm) at 50 °C with a refractive index detector as well as a variable wavelength detector. The instrument was calibrated with linear narrow poly(methyl methacrylate) standards in the range of 0.6 to 47kDa.

Dialysis was carried out in a dialysis kit Thermo Scientific Slide-A-Lyzer Cassette with a molecular weight cut off of 2K. The deionised water was replaced every 6 h and the dialysis was carried out for two days.

End-capping procedure: After the reaction was cooled to room temperature, 0.1 mL of a solution made of 1.00 mg of Pd$_2$(dba)$_3$ and 0.1 mL of 2-(Tributylstannyl)thiophene in 0.5 mL of anhydrous degassed DMF was added and heated for 1 h to 100 °C, then 0.1 mL of a solution made of 1.00 mg of Pd$_2$(dba)$_3$ and 0.1 mL of 2-bromothiophene in 0.5 mL of anhydrous degassed DMF was added and heated to 100 °C.

# 3. Summary of the synthesized materials and their properties

## 3.1 Synthesis of p(gT2)

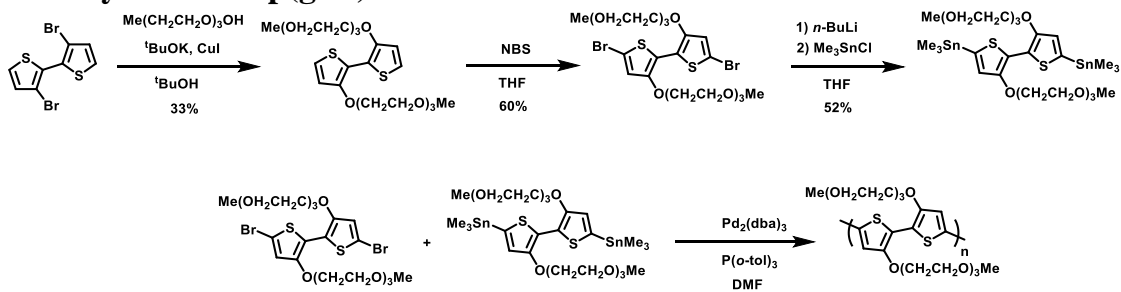

## 3.2 Synthesis of p((DMA)-NDI-gT2)

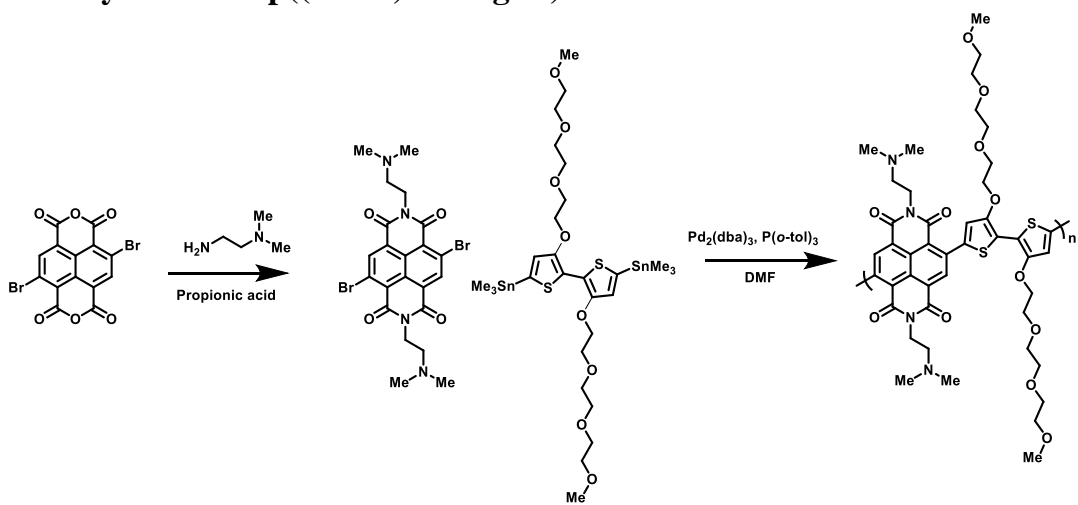

## 3.3 Synthesis of p(ZI-NDI-gT2)

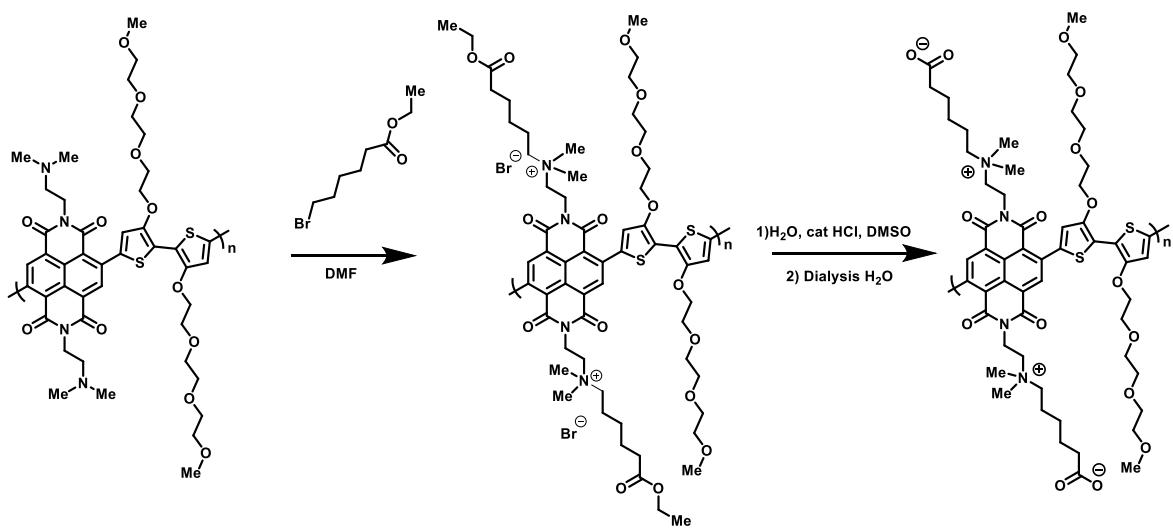

# 4. Monomer synthesis

## 4.1 Synthesis of (DMA)-NDI-Br₂

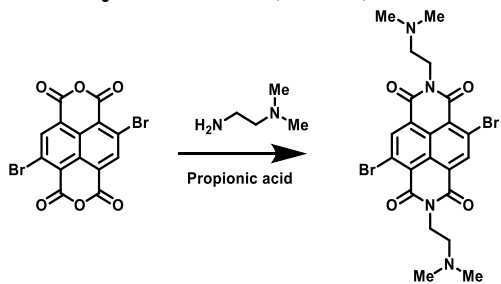

A 150 ml two neck round bottom flask was dried and purged with argon. 2,6-dibromonaphthalene-1,4,5,8-tetracarboxylic dianhydride (512 mg, 1.20 mmol, 1.0 eq.) was suspended in propionic acid (20 ml) and *N,N*-dimethylethane-1,2-diamine (0.23 mg, 2.33 mmol) was added. The reaction mixture was heated to 120°C for 2 h. The reaction was monitored by NMR and after full conversion of the starting material, 100 mL of water and 100 mL of chloroform were added (product is water soluble). The aqueous layer was first washed with chloroform (2 x 100 mL). Then, 200 mL of chloroform was added and a 2 M NaHCO₃ solution was added until a pH value of 8 - 9 was reached. The organic layer was washed with water (3 x 100 mL) and dried over MgSO4. The solvent was removed and the red solid was washed with methanol and acetone. Finally, the solid was washed with hot acetone and dried to obtain 250 mg (0.44 mmol) of an orange solid with a yield of 37 %.

$^1$H-NMR (400 MHz, trifluoroacetic acid -$d_1$) $\delta$: 9.00 (s, 2H), 4.68 (t, 4H), 3.68 (t, 4H), 3.13 (s, 12H) ppm. $^{13}$C-NMR (100 MHz, trifluoroacetic acid-$d_1$) $\delta$: 165.5, 142.5, 132.5, 130.2, 127.0, 126.6, 60.42, 46.3, 38.9 ppm. HRMS (ES-ToF): 565.0090 [M-H]$^+$ (calc. 565.0086).

Figure S1. ¹H NMR spectrum of (DMA)-NDI-Br$_2$ measured in TFA-$d_1$.

Figure S2. ¹³C NMR spectrum of (DMA)-NDI-Br$_2$ measured in TFA-$d_1$.

## 4.2 Synthesis of 3,3'-bisalkoxy(TEG)-2,2'-bithiophene

A 250 mL two neck RBF was dried and purged with argon. 3,3'-Dibromo-2,2'-bithiophene (6.48 g, 20 mmol), triethylene glycol monomethyl ether (11.5 g, 70 mmol), $^t$BuOK (6.73 g, 60 mmol) and CuI (1.52 g, 8.0 mmol) were suspended in anhydrous tert-butanol (100 mL). The reaction mixture was degassed with argon for 15 min and heated to reflux with stirring for 16 h. After the reaction was finished, water was added and the aqueous layer was extracted with ethyl acetate (200 mL). The organic phase was washed with DI water (3 x 100 mL), dried over MgSO$_4$ and concentrated *in vacuo* to afford the crude product as a dark yellow oil. Purification by column chromatography on silica gel with a solvent mixture of ethyl acetate : hexane 3:1 with 1 % of triethylamine afforded the target molecule as a yellow waxy solid (3.2 g, 6.52 mmol, 33% yield).

$^1$H NMR (400 MHz, CDCl$_3$) δ: 7.08 (d, *J* = 5.6 Hz, 2H), 6.85 (d, *J* = 5.6 Hz, 2H), 4.25 (t, *J* = 4.9 Hz, 4H), 3.90 (t, *J* = 5.0 Hz, 4H), 3.78 – 3.75 (m, 4H), 3.69 – 3.66 (m, 4H), 3.66 – 3.64 (m, 4H), 3.55 – 3.53 (m, 4H), 3.37 (s, 6H) ppm. $^{13}$C NMR (100 MHz, CDCl$_3$) δ: 151.9, 122.0, 116.7, 114.9, 72.1, 71.5, 71.1, 70.9, 70.7, 70.2, 59.2 ppm. HRMS (ES-ToF): 491.1774 [M-H$^+$] (calc. C$_{22}$H$_{35}$O$_8$S$_2$ 491.1773).

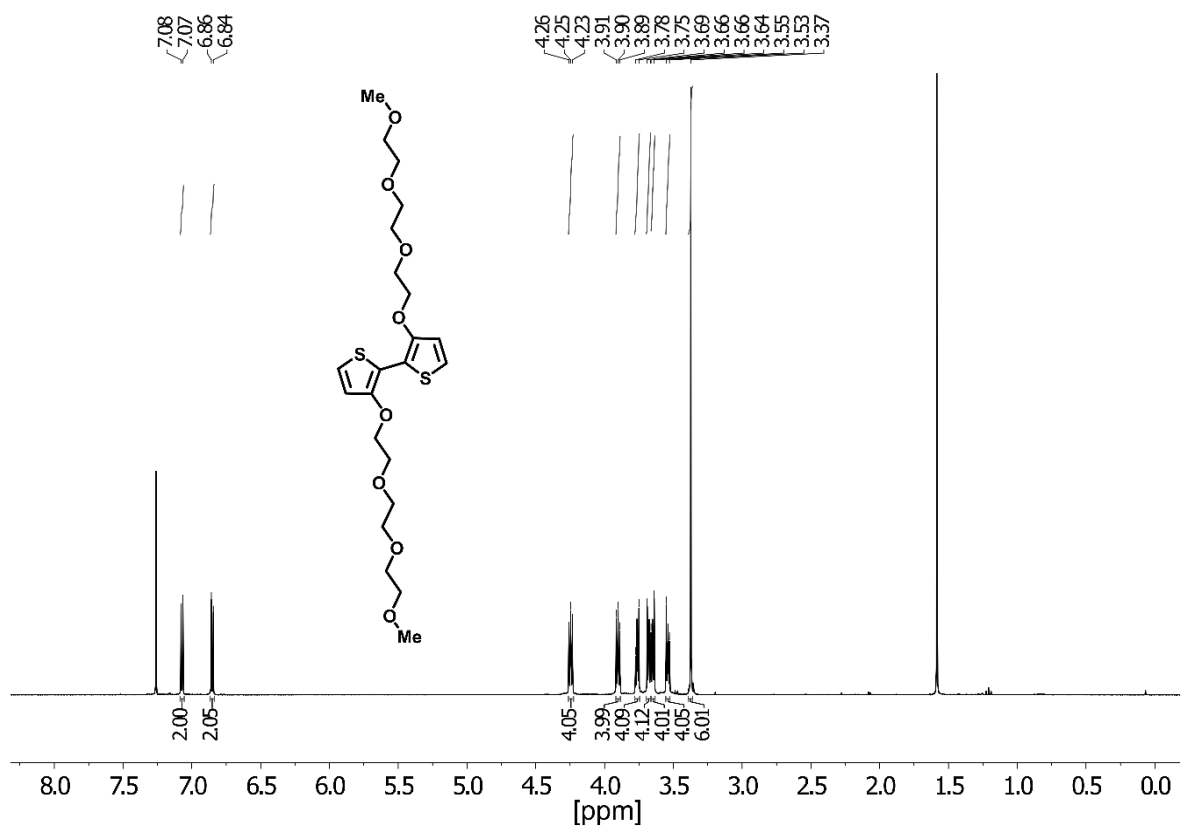

Figure S3. $^1$H NMR spectrum of 3,3'-bisalkoxy(TEG)-2,2'-bithiophene measured in CDCl$_3$.

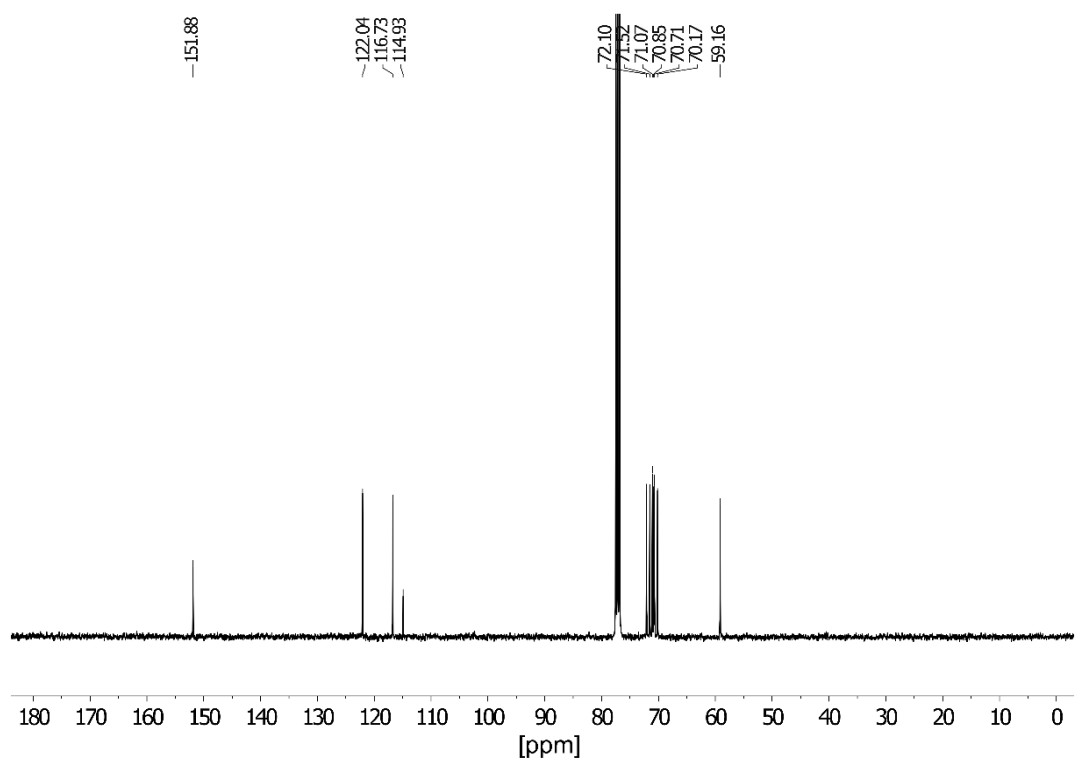

Figure S4. $^{13}$C NMR spectrum of 3,3'-bisalkoxy(TEG)-2,2'-bithiophene measured in CDCl$_3$.

### 4.3  Synthesis of 5,5'-dibromo-3,3'- bisalkoxy(TEG)-2,2'-bithiophene

A protocol by Nielsen *et al.* was followed and slight modifications were carried out.[2] A two neck RBF was dried under vacuum and purged with argon. 3,3'-bisalkoxy(TEG)-2,2'-bithiophene (1.83 g. 2.24 mmol) was dissolved in anhydrous THF (200 mL), degassed with argon and cooled to - 20 °C. NBS (0.83 g, 4.66 mmol) was added in the dark and the reaction mixture was stirred for 10 min (reaction control indicated full conversion after 10 min). The reaction was quenched by the addition of 100 mL of 1 M NaHCO$_3$ aqueous solution, followed by the addition of 150 mL of ethyl acetate. The organic layer was washed with water (3 x 100 mL), dried over MgSO$_4$ and the solvent was removed under reduced pressure. Purification of the crude product was carried out by column chromatography on silica gel with a solvent mixture of hexane : ethyl acetate in the ration of 1 : 1 with 1 % of triethylamine. Finally, the product was recrystallised from diethyl ether/hexane to afford the product as a yellow solid with a yield of 60 %. (1.45 g, 2.24mmol).

$^1$H NMR (400 MHz, CDCl$_3$) δ: 6.85 (s, 2H), 4.21 – 4.18 (m, 4H), 3.88 – 3.85 (m, 4H), 3.75 – 3.73 (m, 4H), 3.70 – 3.68 (m, 4H), 3.68 – 3.65 (m, 4H), 3.57 – 3.54 (m, 4H), 3.38 (s, 6H) ppm. $^{13}$C NMR (100 MHz, CDCl$_3$) δ: 150.3, 119.8, 116.0, 111.2, 72.1, 71.8, 71.1, 70.9, 70.8, 70.3, 59.2 ppm. HRMS (ES-ToF): 646.9964 [M-H$^+$] (calc. C$_{22}$H$_{33}$Br$_2$O$_8$S$_2$ 646.9983).

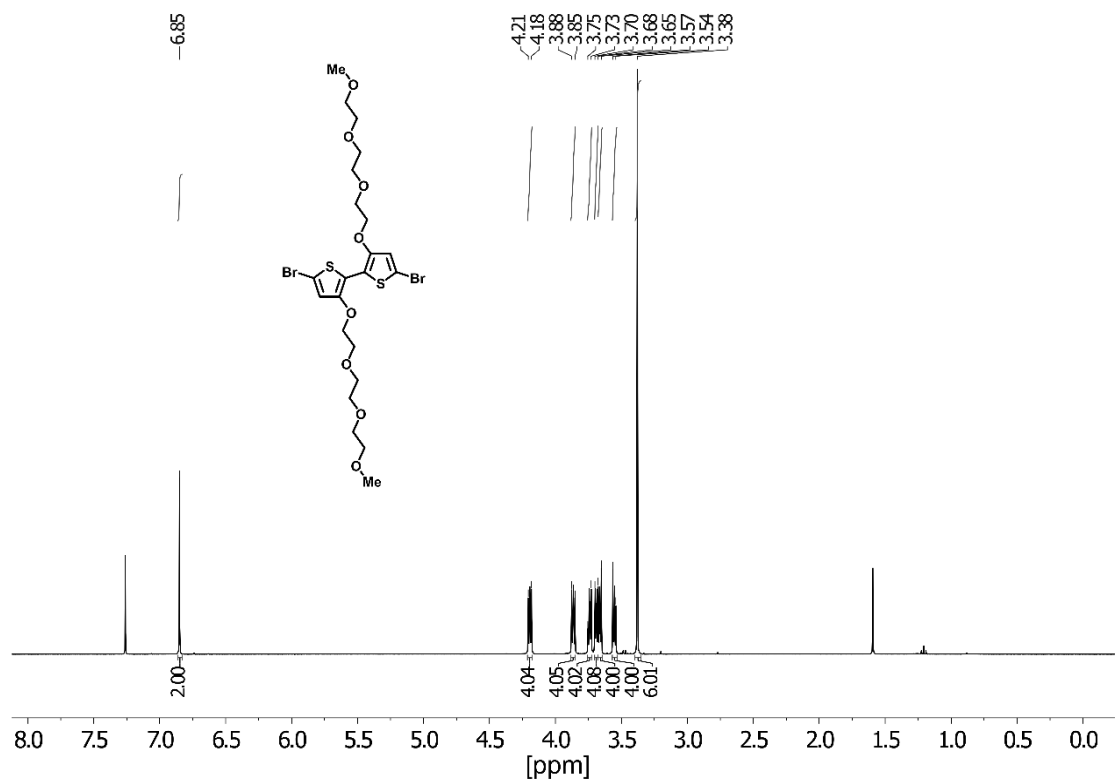

Figure S5. $^1$H NMR spectrum of 5,5'-dibromo-3,3'- bisalkoxy(TEG)-2,2'-bithiophene measured in CDCl$_3$.

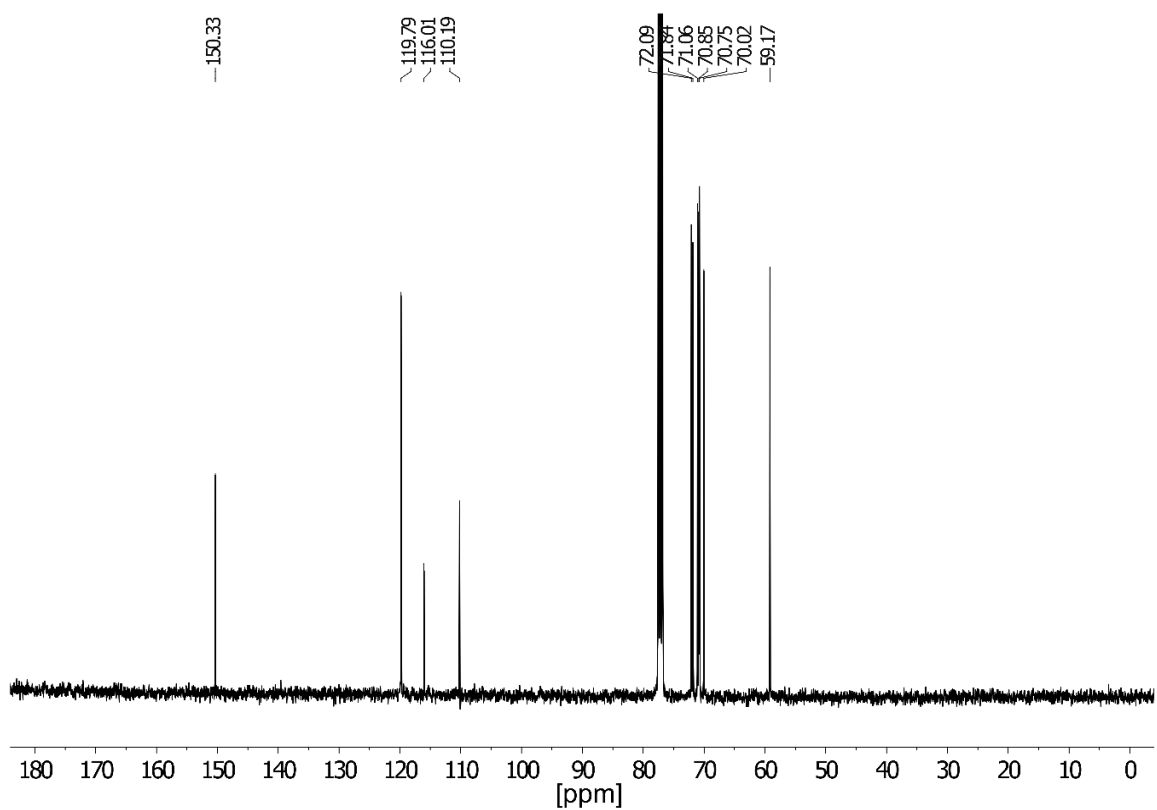

Figure S6. $^{13}$C NMR spectrum of 5,5'-dibromo-3,3'- bisalkoxy(TEG)-2,2'-bithiophene measured in CDCl$_3$.

## 4.4 Synthesis of (3,3'- bisalkoxy(TEG)-[2,2'-bithiophene]-5,5'-diyl)bis(trimethylstannane)

A 250 mL two neck RBF was dried and purged with argon. 1.01 g of 5,5'-dibromo-3,3'- bisalkoxy(TEG)-2,2'-bithiophene (1.56 mmol, 1.0 eq.) was dissolved in 150 mL of anhydrous THF. The reaction mixture was cooled to -78 °C and 2.5 mL of *n*-BuLi (2.5 M in hexane, 6.23 mmol, 4.0 eq.) was added slowly. A yellow solution was formed which was stirred at -78 °C for 3 h. Then, 7.8 mL of trimethyltin chloride (1 M in hexane, 7.8 mmol, 5.0 eq.) was added and the reaction mixture was warmed to room temperature. 150 mL of diethyl ether was added and the organic phase was washed with water (3 x 100 mL) and dried over $Na_2SO_4$. The solvent was removed and the obtained solid was dissolved in 100 mL of acetonitrile which was washed with hexane (3 x 100 mL). The solvent was removed and the product was recrystallized from diethyl ether to obtain the product as yellow needles with a yield of 64 % (810 mg, 0.99 mmol).

$^1$H NMR (400 MHz, acetone-*d$_6$*) δ: 8.99 (s, 2H), 4.28 – 4.26 (m, 4H), 3.88 – 3.85 (m, 4H), 3.71 – 3.69 (m, 4H), 3.62 – 3.56 (m, 8H), 3.46 – 3.44 (m, 4H), 3.27 (s, 6H), 0.37 (s, 18 H) ppm. $^{13}$C NMR (100 MHz, acetone-*d$_6$*): 154.9, 134.9, 125.3, 121.1, 72.8, 72.4, 71.8, 71.3, 71.0, 59.0, 8.24 ppm. HRMS (ES-ToF): 819.1069 [M-H+] (calc. 819.1084).

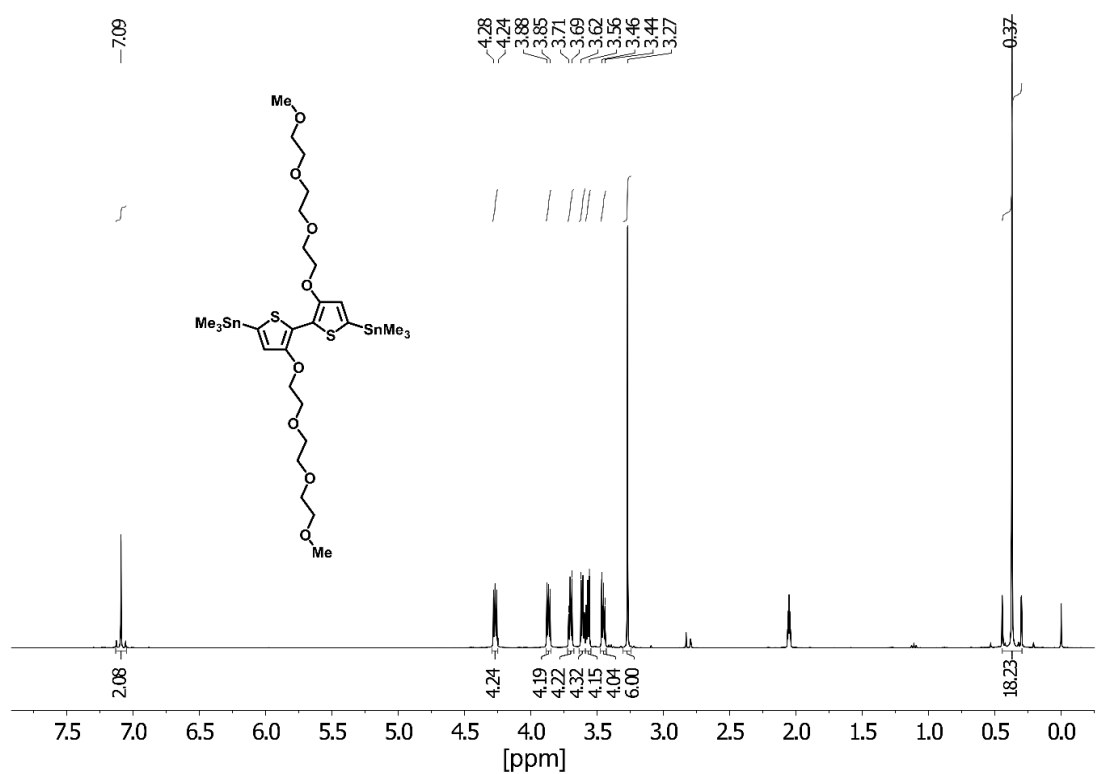

Figure S7. $^1$H NMR spectrum of (3,3'- bisalkoxy(TEG)-[2,2'-bithiophene]-5,5'-diyl)bis(trimethylstannane) in acetone-*d$_6$*.

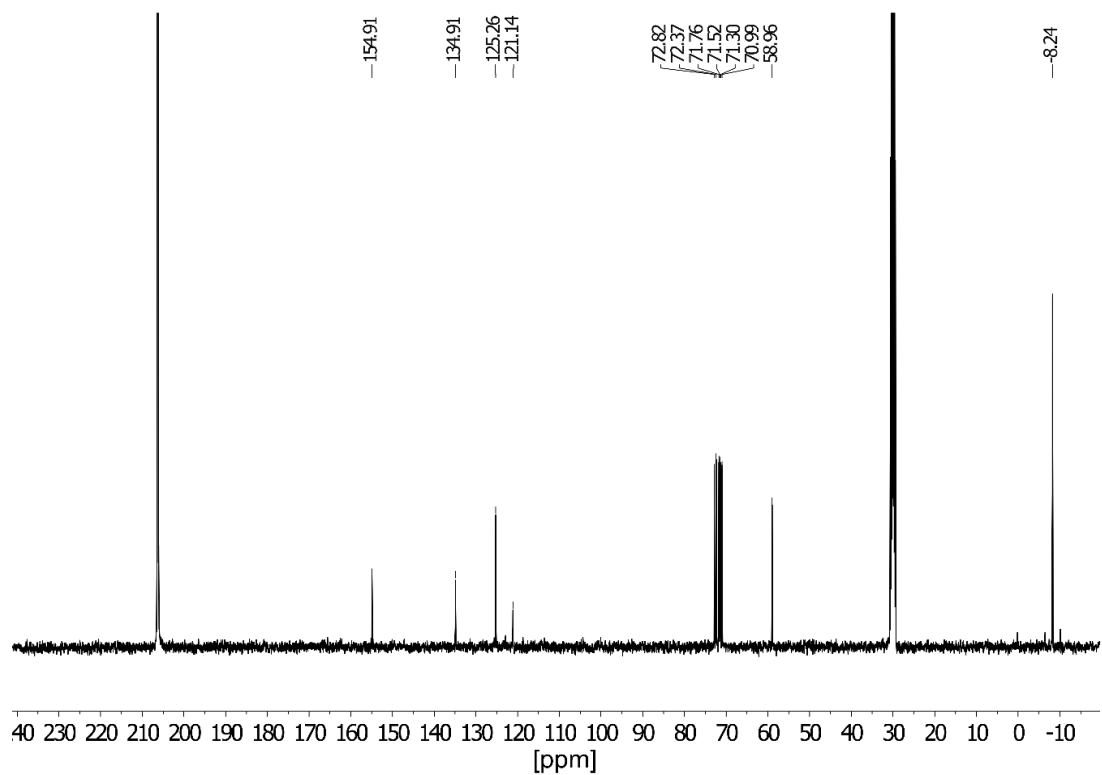

Figure S8. $^{13}$C NMR spectra of (3,3'- bisalkoxy(TEG)-[2,2'-bithiophene]-5,5'-diyl)bis(trimethylstannane) in acetone-$d_6$.

# 5. Polymer synthesis

## 5.1 Synthesis of p(gT2)

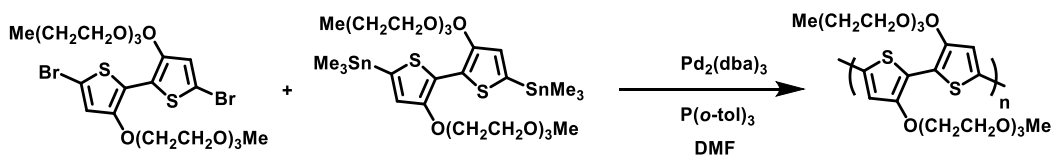

In a dried 5.0 mL microwave vial, 79.4 mg of (3,3'- bisalkoxy(TEG)-[2,2'-bithiophene]-5,5'-diyl)bis(trimethylstannane) (97.3 µmol) and 63.1 mg of 5,5'-dibromo-3,3'- bisalkoxy(TEG)-2,2'-bithiophene (97.3 µmol) were dissolved in 3.0 mL of anhydrous, degassed DMF. $Pd_2(dba)_3$ (1.78 mg, 1.95 µmol) and P(o-tol)$_3$ (2.37 mg, 7.78 µmol) were added and the vial was sealed and heated to 100 °C for 16 h. After the polymerization has finished, the end-capping procedure was carried out. Then, the reaction mixture was cooled to room temperature and precipitated in methanol. A blue solid was formed which was filtered into a glass fibre-thimble and Soxhlet extraction was carried out with methanol, ethyl acetate, acetone, hexane, and chloroform. The polymer dissolved in hot chloroform. Finally, the polymer was dissolved in a minimum amount of chloroform and precipitated in methanol. The collect solid was filtered and dried under high vacuum. A blue solid was obtained with a yield of 74 % (70 mg, 71.6 µmol).

GPC (DMF, 50 °C) $M_n$ = 36 kDa, $M_w$ = 60 kDa (including the first fraction $M_n$ = 56 kDa, $M_w$ = 306 kDa). $^1$H NMR (CDCl$_3$, 400 MHz) δ: 6.96 (br s, 2 H), 4.35 (br s, 4 H), 4.01 – 3.92 (m, 4 H), 3.81 – 3.79 (m, 4 H), 3.72 – 3.70 (m, 4 H), 3.66 – 3.63 (m, 4 H), 3.53 – 3.50 (m, 4 H), 3.34 (s, 6 H) ppm.

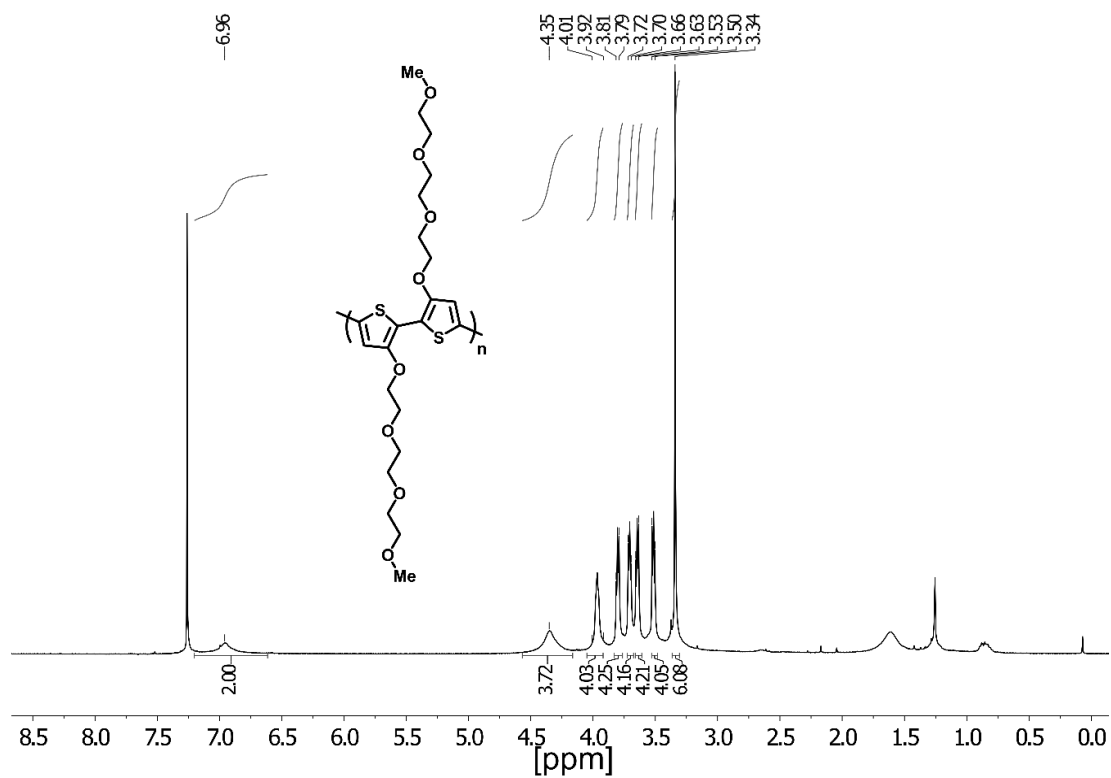

Figure S9. $^1$H NMR spectrum of p(gT2) in CDCl$_3$ at 25 °C.

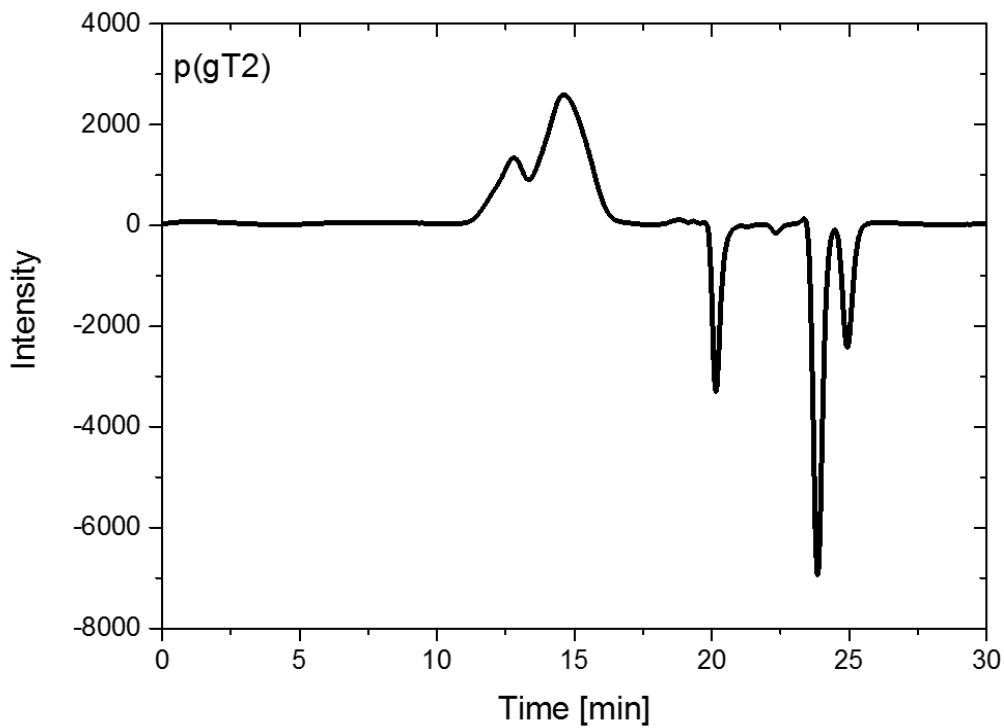

Figure S10: GPC measurements of p(gT2) in DMF (bimoduale eluation).

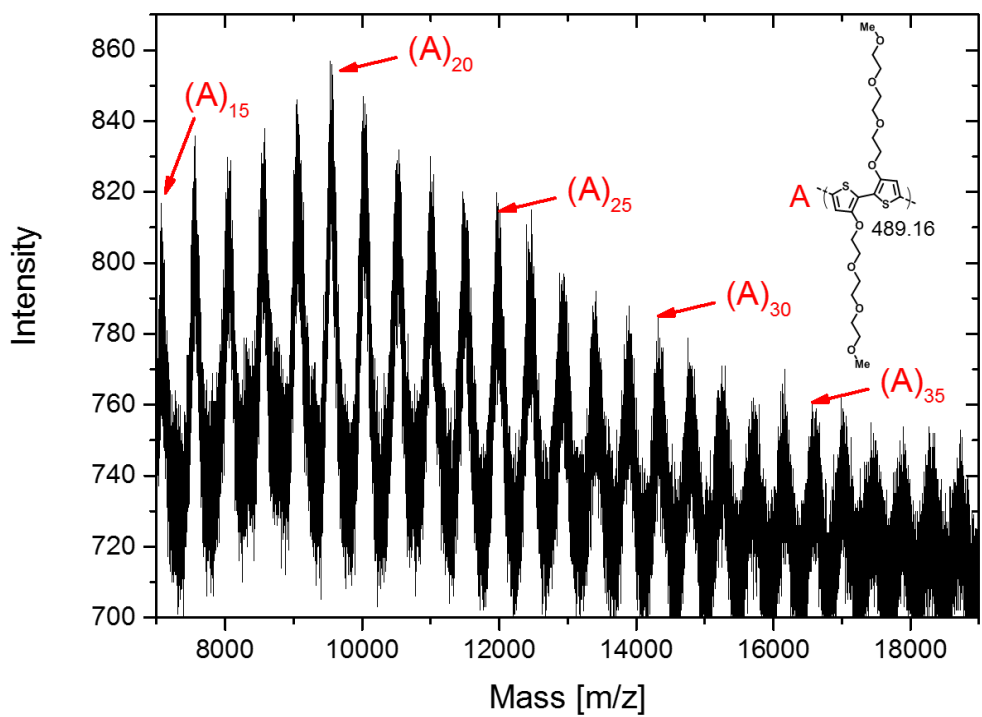

Figure S11. MALDI-ToF spectrum of p(gT2), measured in positive linear mode with DCTB as the matrix.

## 5.2 Synthesis of p((DMA)-NDI-gT2)

In a 5.0 mL microwave vial, 38.76 mg of (DMA)-NDI-Br$_2$ (68.45 µmol, 1.0 eq.), 55.86 mg of (3,3'-bisalkoxy(TEG)-[2,2'-bithiophene]-5,5'-diyl)bis(trimethylstannane) (68.45 µmol, 1.0 eq.), 1.4 mg of Pd$_2$(dba)$_3$ (1.52 µmol, 2 mol%) and 1.87 mg of P(o-tol)$_3$ (6.12 µmol, 8 mol%) were dissolved in 1.5 mL of anhydrous, degassed DMF and the vial was heated to 85 °C for 16 h. The color changed from yellow to dark green. After the polymerization has finished, the end-capping procedure was carried out. Then, the reaction mixture was cooled to room temperature and the reaction mixture was precipitated in ethyl acetate. The precipitate was filtered and Soxhlet extraction was carried out with ethyl acetate, methanol, acetone, hexane and chloroform. The polymer was soluble in hot chloroform. Polymer p((DMA)-NDI-gT2) was obtained as a green solid with a yield of 79 % (48.4 mg, 54.1 µmol).

GPC (CHCl$_3$, 50 °C) M$_n$ = 5.3 kDa, M$_w$ = 8.9 kDa. $^1$H-NMR (400 MHz, CHCl$_3$) δ: 8.82 (br s, 2H), 7.26 (br s, 4H), 4.42 (br s, 4H), 4.35 (br s, 4H), 4.26 (br s, 4H), 3.91 – 383 (m, 4H), 3.83 – 3.75 (m, 4H), 3.75 – 3.64 (m, 4H), 3.64 – 3.54 (m, 4H), 3.54 – 3.45 (m, 4H), 3.39 – 3.28 (m, 6H), 2.69 – 2.58 (m, 4H), 2.40 – 2.22 (m, 12H) ppm.

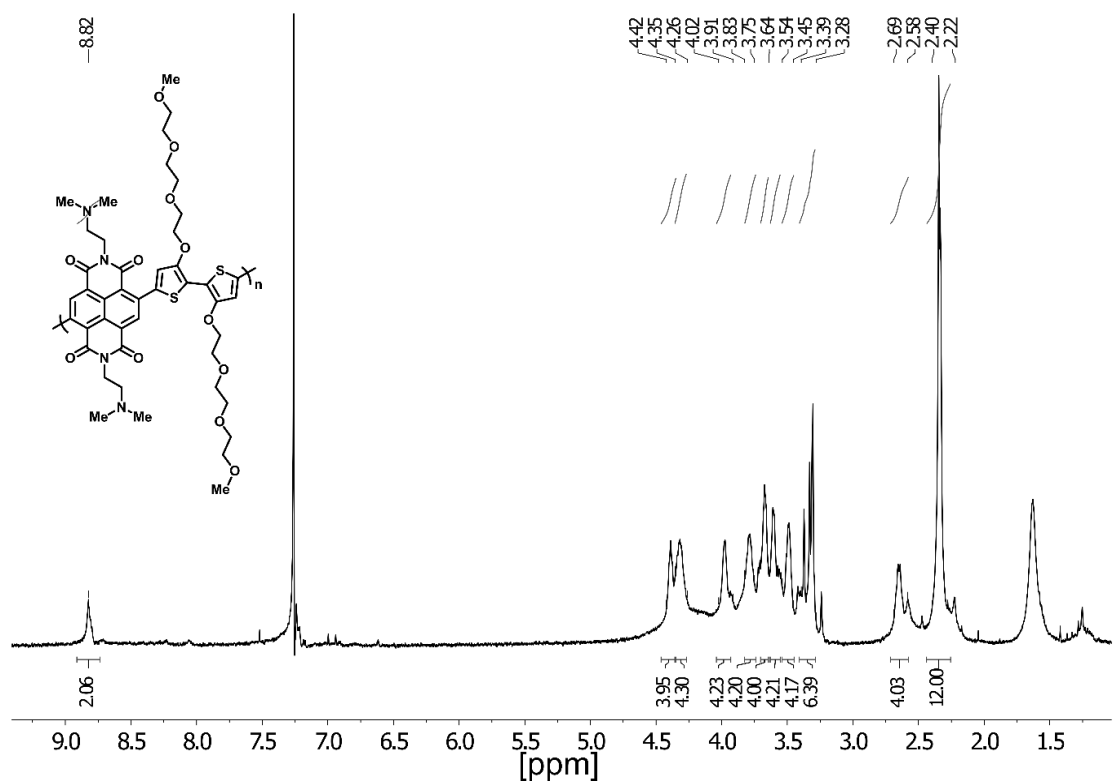

Figure S12. $^1$H NMR spectrum of p((DMA)-NDI-gT2) in CDCl$_3$ at 25 °C.

S10

Figure S13. MALDI-ToF spectrum of p((DMA)-NDI-gT2).

## 5.3 Synthesis of p((DMA-Br)-NDI-gT2)

A 5.0 mL microwave vial was dried and purged with argon, 14.3 mg p((DMA)-NDI-gT2) (15.9 µmol) was suspended in 4.00 ml anhydrous DMF and 0.5 ml ethyl-6-bromohexanoate (2.8 mmol) was added. The reaction mixture was heated to 120°C for 1.5 h and a green solution was formed. The reaction mixture was cooled to room temperature and the solvent was removed under reduced pressure. The green solid was suspended in 3 mL of methanol and precipitated in acetone followed by the addition of hexane. The solution was filtered and the green solid was washed with chloroform and acetone. Finally, the polymer was dried under high vacuum for 16 h. 20.6 mg (15.4 µmol) of the polymer was obtained as a green solid with a yield 97%.

GPC (DMF, 50 °C) $M_n$ = 38 kDa, $M_w$ = 60 kDa. $^1$H-NMR (500 MHz, DMSO-$d_6$) δ: 8.64 – 8.57 (m, 2H), 7.81 – 7.48 (m, 4H), 4.49 – 4.24 (m, 8H, signals overlapping), 4.07 – 3.98 (m, 4H, COCH$_2$), 3.98 – 3.89 (m, 4H), 3.68 – 3.62 (m, 4H), 3.57 – 3.50 (m, 8H), 3.47 – 3.40 (m, 8H), 3.22 – 3.11 (m, 12H), 2.95 – 2.86 (m, 4H), 3.98 – 3.89 (m, 4H), 2.37 – 2.29 (m, 4H), 1.80 – 1.70 (m, 4H), 1.62 – 1.54 (m, 4H), 1.37 – 1.27 (m, 4H), 1.18 – 1.13 (m, 6H) ppm.

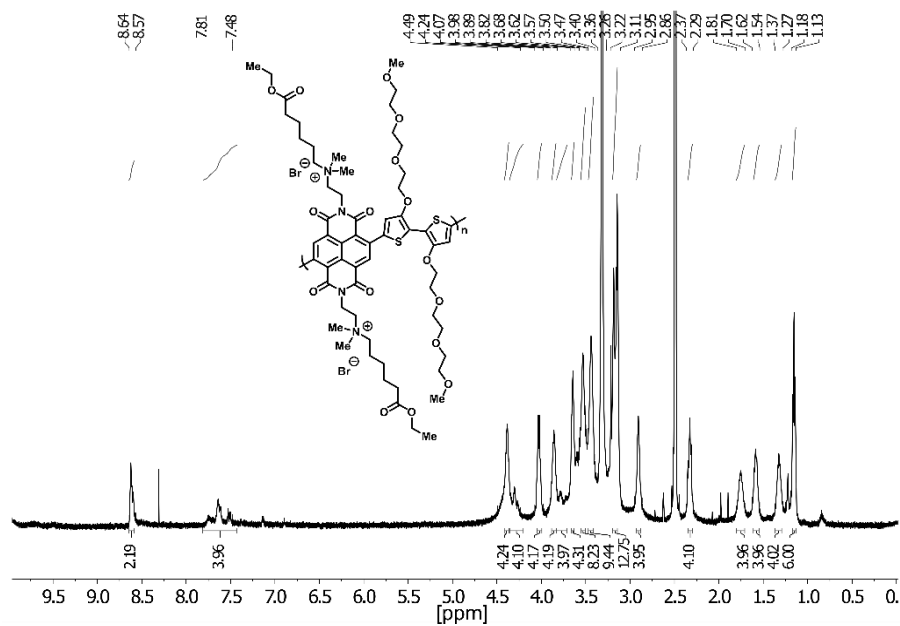

Figure S14. $^1$H NMR spectrum of p((DMA-Br)-NDI-gT2) in DMSO-$d_6$ at 22 °C.

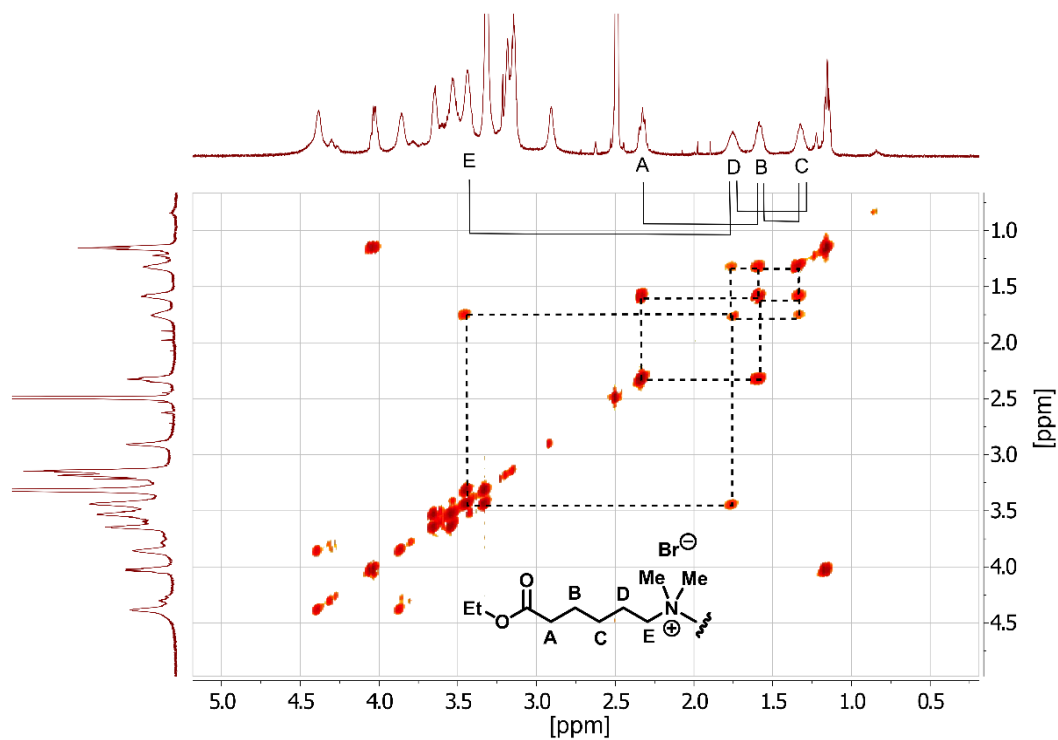

Figure S15. 2D COSY spectrum of p((DMA-Br)-NDI-gT2) in DMSO-$d_6$ at 25 °C, proton couplings of the alkylation are highlighted.

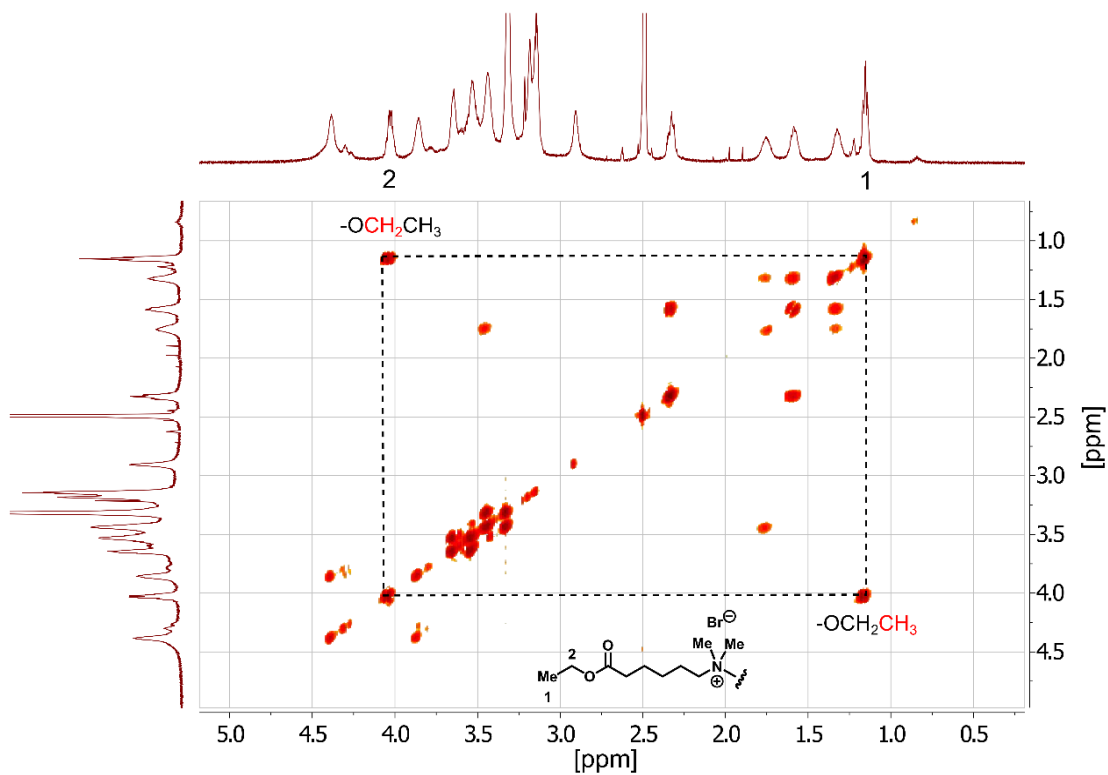

Figure S16. 2D COSY spectrum of p((DMA-Br)-NDI-gT2) in DMSO-$d_6$ at 25 °C, proton couplings of ester group are highlighted.

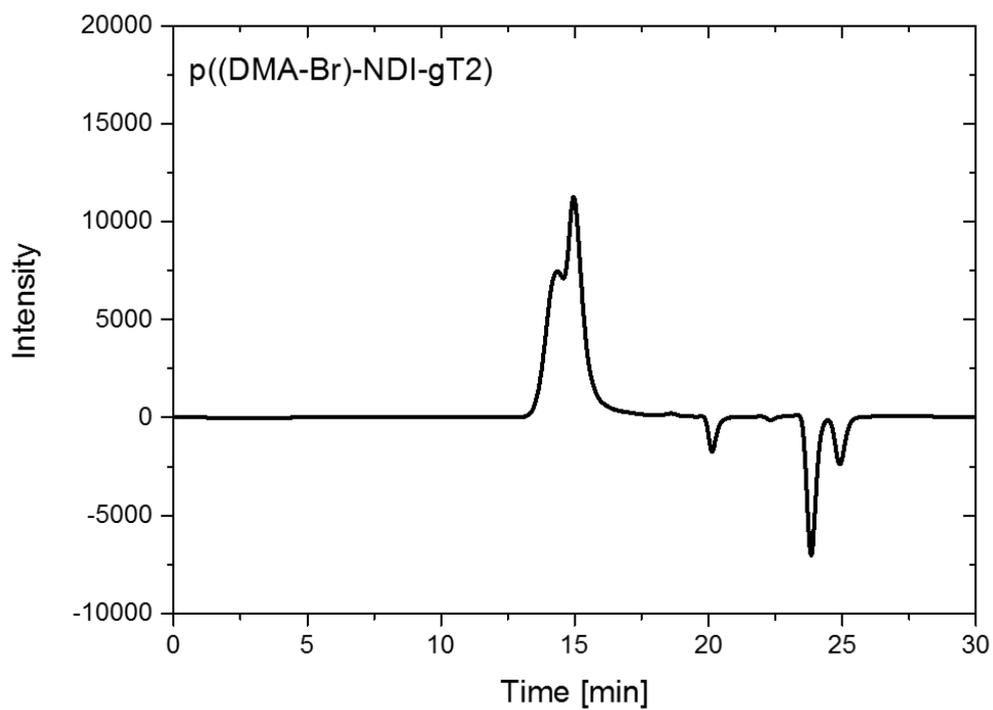

Figure S17. GPC measurements of p((DMO-Br)-NDI-gT2) in DMF (bimoduale eluation).

## 5.4 Synthesis of p(ZI-NDI-gT2)

In a 7.0 mL microwave, 10 mg of p((DMA-Br)-NDI-gT2) (8.46 µmol, 1.00 eq.) was dissolved in 2 mL of DMSO and 2 mL of water. 0.1 mL of conc. HCl was added and the reaction mixture was heated to 75 °C for 16 h. Then, the reaction mixture was cooled to room temperature and the solution was transferred into a dialyses kit (molecular weight cut off 2kDa) and the dialysis kit was stirred in DI water for 2 days, exchanging the water every 6 h. Finally, solvent was removed and the polymer was dried at 60 °C for 16 h. 9.2 mg (8.19 µmol) of a green polymer was obtained with a yield of 97 %.

GPC (DMF, 50 °C) $M_n$ = 24 kDa, $M_w$ = 53 kDa. $^1$H-NMR (500 MHz, DMSO-$d_6$) δ: 8.67 – 8.55 (m, 2H), 7.73 – 7.45 (m, 4H), 4.48 – 4.30 (m, 8H, signals overlapping), 3.61 – 3.56 (m, 8H, signals overlapping), 3.55 – 3.52 (m, 4H), 3.46 – 3.40 (m, 8H), 3.40 – 3.24 (m, 4H, overlap with water peak), 3.23 – 3.12 (m, 12H), 2.30 – 2.21 (m, 4H), 1.81 – 1.70 (m, 4H), 1.62 – 1.54 (m, 4H), 1.37 – 1.28 (m, 4H) ppm.

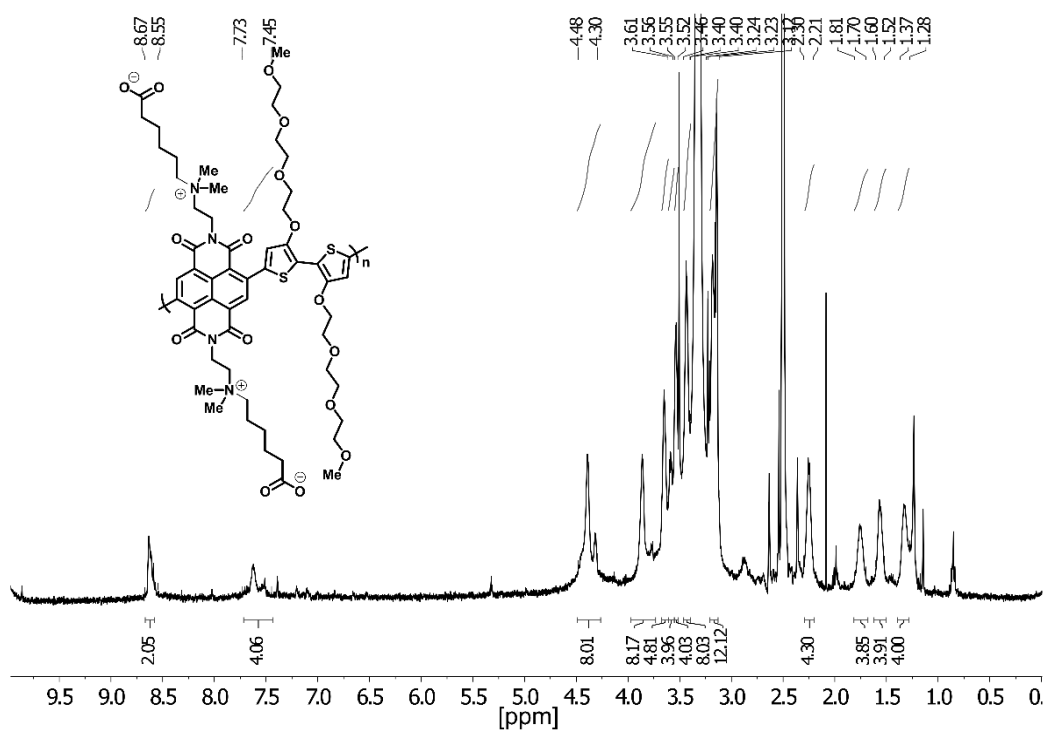

Figure S18. $^1$H NMR spectrum of p(ZI-NDI-gT2) in DMSO-$d_6$.

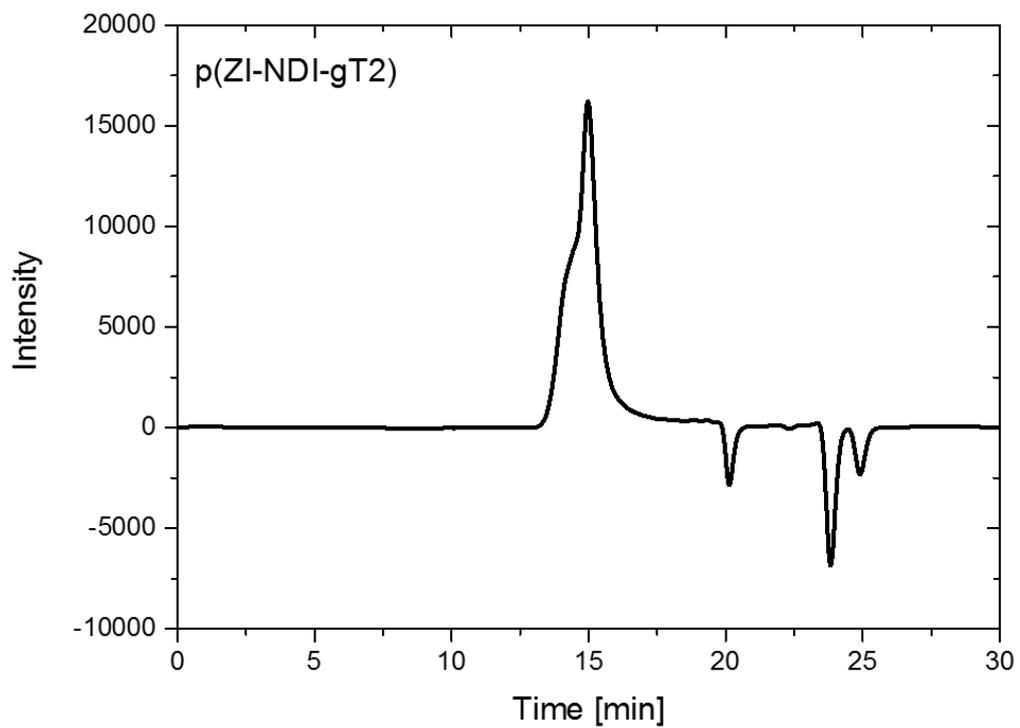

Figure S19. GPC of p(ZI-NDI-gT2) in DMF (bimoduale eluation).

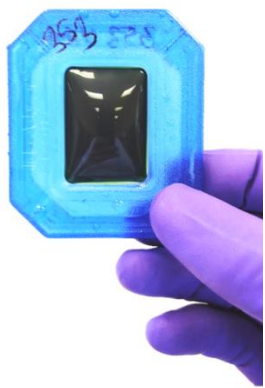

Figure S20. Dialysis of p(ZI-NDI-gT2) in DI water.

## 5.5 Synthesis of p(g7NDI-gT2)

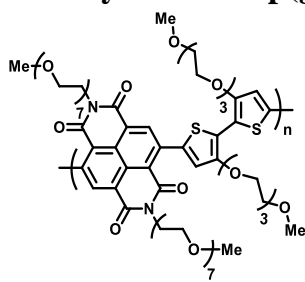

In a 7.0 mL microwave vial, 72.25 mg of g7NDI-Br$_2$ (67.6 μmol, 1.0 eq.) and 55.17 mg of (3,3'-bisalkoxy(TEG)-[2,2'-bithiophene]-5,5'-diyl)bis(trimethylstannane) (67.6 μmol, 1.0 eq.) were dissolved in 2.0 mL of anhydrous, degassed chlorobenzene. 1.36 mg of Pd$_2$(dba)$_3$ (1.35 μmol, 2 mol%) and 1.78 mg of P(o-tol)$_3$ (5.4 μmol, 8 mol%) were added and the vial was heated to 135 °C for 16 h. After the polymerization has finished, the end-capping procedure was carried out. Then, the reaction mixture was cooled to room temperature and the dark green reaction mixture was precipitated in ethyl acetate followed by addition of hexane. Soxhlet extraction was carried out with hexane, ethyl acetate, MeOH, acetone, THF and chloroform. The polymer was soluble in hot chloroform. Polymer p(g7NDI-gT2) was obtained as a dark green solid with a yield of 76 % (48 mg, 0.05 mmol).

GPC (CHCl$_3$:DMF; 5:1, 50 °C) M$_n$ = 14 kDa, M$_w$ = 26 kDa. $^1$H NMR (400 MHz, CDCl$_3$) δ: 8.82 (s, 2H), 7.31 – 7.16 (m, 4H), 4.44 – 4.34 (m, 4H), 4.34 – 4.25 (m, 4H), 4.02 – 3.93 (m, 4H), 3.89 – 3.75 (m, 4H), 7.31 – 7.16 (m, 4H), 3.70 – 3.52 (m, 60H), 3.52 – 3.49 (m, 4H), 3.36 (br s, 12H).

CV (PESA) = 5.1 eV

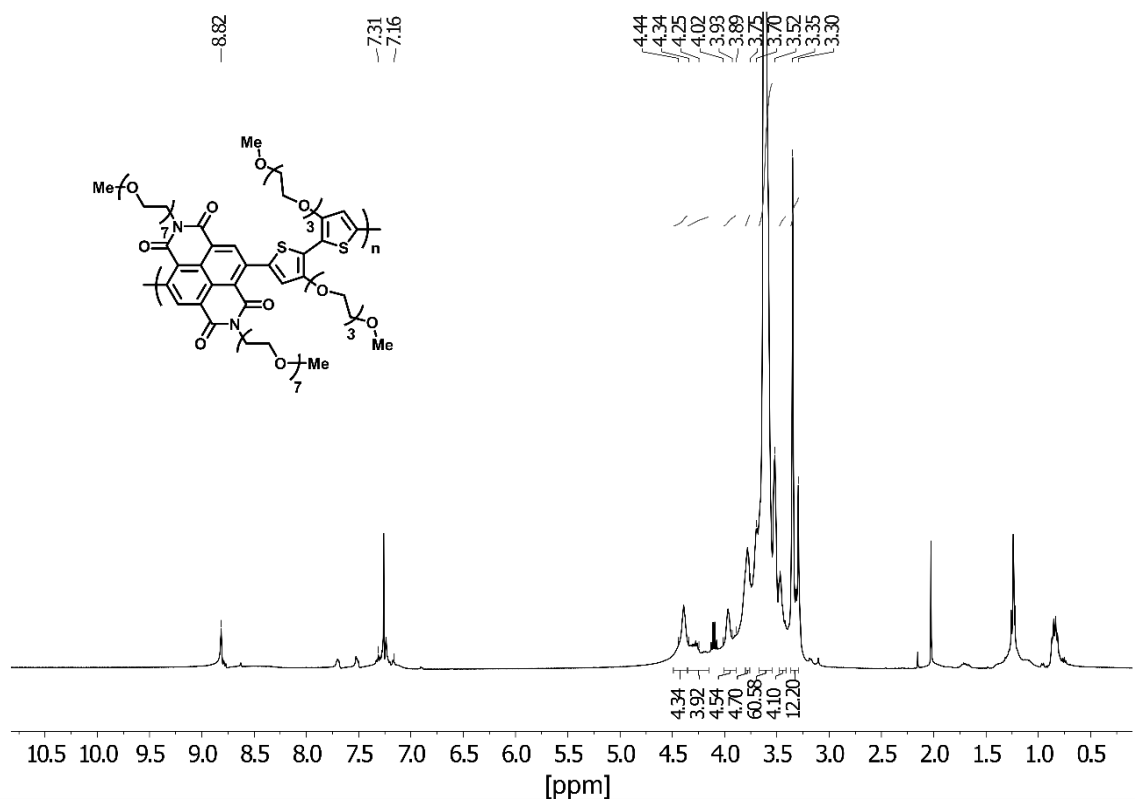

Figure S21. $^1$H NMR spectrum of polymer p(g7NDI-gT2) in CDCl$_3$ at 22 °C, aromatic signals were not integrated due to the overlap with the solvent peak.

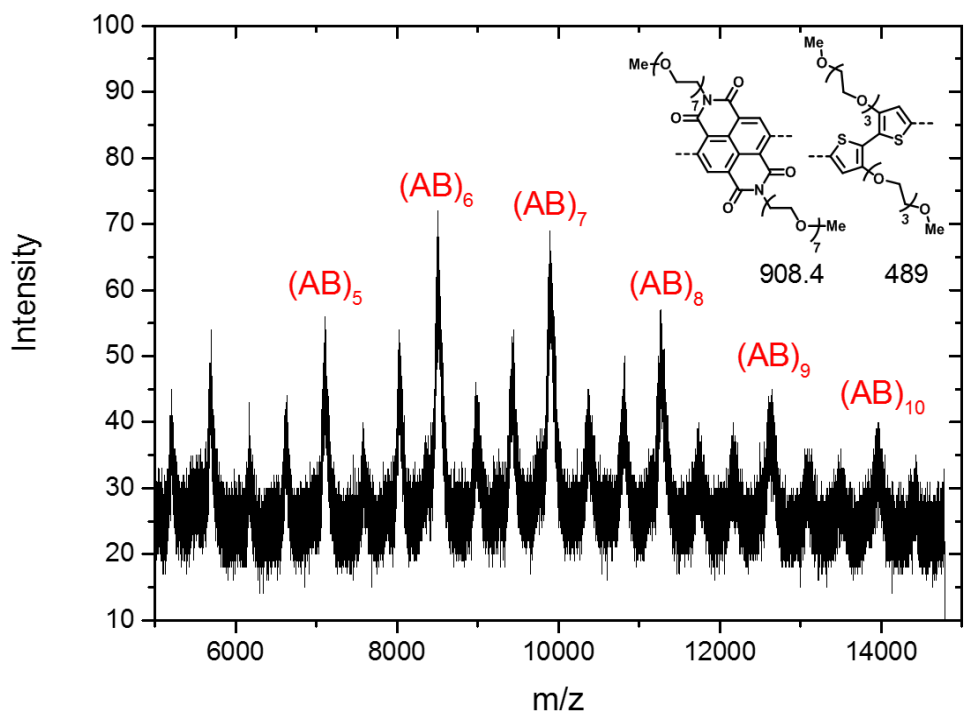

Figure S22. MALDI-ToF mass spectrum of p(g7NDI-gT2)

# 6. CV measurements in organic electrolytes

Thin film CV measurements reported in this section and in Figure 2b in the main text were carried out on FTO coated glass substrates with a 0.1 M TBAPF$_6$ acetonitrile solution as the supporting electrolyte. Potentials are reported vs Ag/AgCl and the scan rate of the measurement was 100 mV/s. The oxidation of ferrocene (solution) was used as the reference to calculate IP and EA values ($E_{1/2}$ (Fc/Fc$^+$) = 0.46 V vs Ag/AgCl)

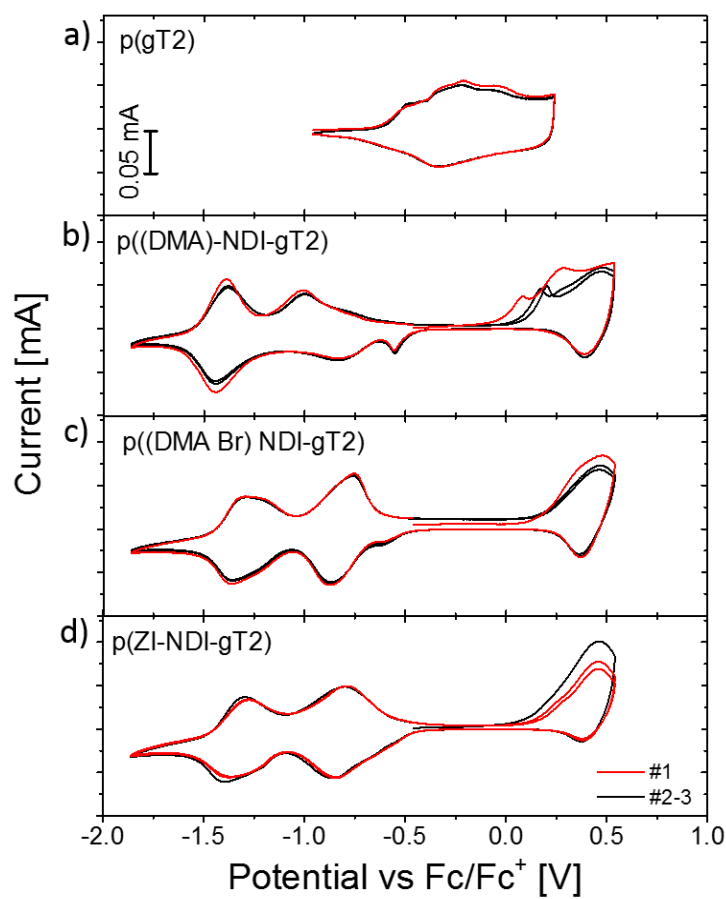

Figure S23: CV measurements of a) p(gT2), b) p((DMA)-NDI-gT2), c) p((DMA-Br)-NDI-gT2), d) p(ZI-NDI-gT2) with a degassed 0.1 M NBu$_4$PF$_6$ acetonitrile solution as the supportive electrolyte.

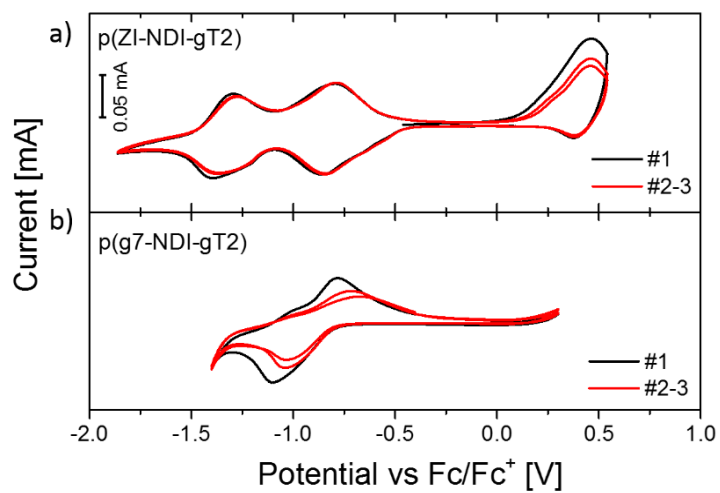

Figure S24: CV of a) p(ZI-gNDI-gT2) and b) p(g7-NDI-gT2) with a degassed 0.1 M NBu$_4$PF$_6$ acetonitrile solution as the supportive electrolyte. Polymer p(g7-NDI-gT2) showed limited redox-stability where applying higher voltages results in irreversible charging.

# 7. UV Vis measurements

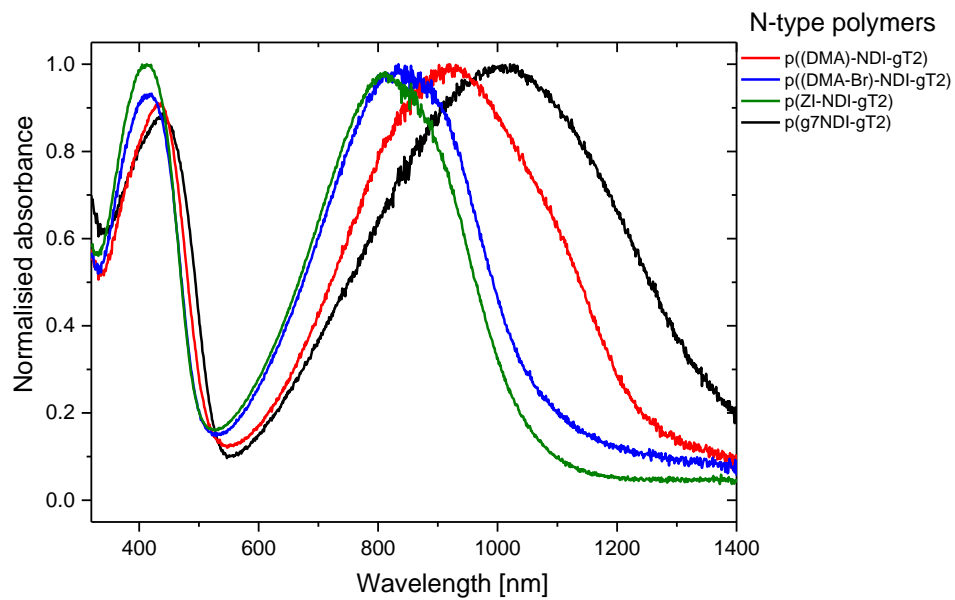

Figure S25. Normalized UV-Vis spectra of the n-type copolymers prepared by spin coating or doctor blade coating on glass substrates.

# 8. Electrochemical and spectroelectrochemical measurements

## 8.1 Sample preparation:

Spectroelectrochemical measurements:

The preparation of samples presented in the main text was carried out according to the following procedure: fluorine doped tin oxide (FTO TEC15) or indium doped tin oxide (ITO) coated glass substrates were cleaned with soap, deionized water acetone and isopropanol before undergoing a heating step at 450°C for 30 minutes. The polymers were dissolved in organic solvents (chloroform for p(gT2) and p(g7NDI-gT2) (5 mg/mL) and DMSO for p(ZI-NDI-gT2) (5 mg/mL)) and deposited from solution onto the substrates via spin coating, drop casting or doctor blading.

p(gT2) and p(g7NDI-gT2) were deposited at room temperature, while for p(ZI-NDI-gT2) the deposition was performed on a hot plate and followed by an additional heating step in order to remove residuals of solvent in the film. Temperatures in the range between 120 and 160°C were used for the deposition and post-processing of the film and yielded similar results in terms of electrochemical characterization. Araldite rapid epoxy glue was added whenever it was needed to minimize exposure of FTO/ITO/Au surface to the electrolyte. The measurement of thin film thickness for the polymers was carried out by calibrating against a reference value of absorbance of a samples with known thickness (thickness was measured by using a Dektak profilometer). The thickness of thin films was calculated from optical absorbance of the samples. Films that were thicker than 200 nm were measured with the profilometer.

Volumetric capacity measurements:

Thick films (>400 nm) were deposited on gold coated glass substrates (with a thin interlayer of parylene to improve adhesion of the gold layer) by either drop casting or blade coating. The area of the devices was defined by removing parts of the polymer and the area was measured with a profilometer. Film thickness was then measuremed by taking the average of at least 6 measurements on each films using a profilometer (in dry-state) before electrochemical characterization. The active area of the samples in this case was in the order of 0.1 cm$^2$.

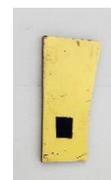

Drop cast film of p(gT2) on gold substrates with a defined area.

Gravimetric capacity measurements:

For gravimetric measurements of p(gT2), a dilute solution of the polymer was prepared in chloroform (2,2 mg/mL) and 10 µL of this solution was deposited by drop-cast on gold coated substrates (with parylene interlayer). For p(ZI-NDI-gT2), a solution of 3 mg/mL was prepared and the the solution was deposited whilst heating the substrate to 120 °C, followed by an annealing step at 140 °C for 30 mins.

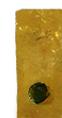

Drop cast film on gold substrates

## 8.2  Electrochemical measurements in aqueous electrolytes

Electrochemical characterization presented in Figure 4 in the main text was carried out using an Ivium CompactStat potentiostat in an electrochemical cell filled with 0.1 M NaCl aqueous electrolyte (unless stated otherwise), using the polymer film as the working electrode of the cell, a Ag/AgCl 3 M NaCl reference electrode and a platinum-mesh counter electrode (active area 25 x 35 mm). The electrolyte was flushed with argon for at least 15 minutes before the beginning of measurements to avoid faradaic side reactions with molecular oxygen during reduction of the n-type polymers. Galvanostatic measurements were performed by applying a sequence of charging and discharging constant current levels to the sample and monitoring the voltage of the sample (for measurements of single electrodes) or of the cell (for measurements on the two-electrode cell). The charging and the discharging currents were of same magnitude in all cases. Multiple consecutive cycles were performed to evaluate the rate capabilities of the electrodes and of the two-electrode cell for each value of applied current and the $2^{nd}$ cycle was used to calculate the specific capacity illustrated in Figure 5. The value of C rate for each Galvanostatic measurement was calculated on the basis of the set current density and the value of reversible charge density obtained from the measurement using the slowest rate performed on the electrode. Regarding the cyclic voltammetry characterization, the two-electrode cell was scanned to a negative potential (between -0.8 V and -1 V) to reset the neutrality of the p-type polymer between scans presented in Figure 6c. This procedure was followed to avoid distortion in the CV profiles due to charge retention unbalance between the cathode and anode electrodes. This effect and its consequences are discussed in detail in Figure S52.

## 8.3  Spectroelectrochemical measurements in aqueous electrolytes

The electrochemical cell was a quartz cuvette with transparent windows which enabled simultaneous optical spectra acquisition through a UV-vis spectrometer (OceanOptics USB 2000+) which collected the transmitted light through the sample from a tungsten lamp used as probe light source. Spectroelectrochemical measurements on a two-electrode cell including both p-type and an n-type polymer films were performed in a similar way, using a 2 electrode configuration and by monitoring the optical absorption of both films in series.

# 9. Quantum chemical calculations

Molecules in their neural, singly and doubly reduced (for n-type) or oxidized (for p-type) states were optimized at the DFT level (B3LYP/6-31g(d,p)). TD-DFT calculations were then performed on the optimized geometries (B3LYP/6-31g(d,p)) to obtain absorption spectra. All quantum chemical calculations were done using Gaussian16.[3]

In order to reproduce experimental data on p(gT2), oligomers (gT2)$_n$ with n=1, 3 and 6 were calculated using above described procedure. For neutral systems, the lowest energy absorption peak position red-shifts with increasing number of monomer units from 328 nm for monomer, 583 nm for trimer to 763 nm for hexamer. These spectra are presented in Figure S26. The experimental value is around 645 nm as shown in the main text. On this basis, we chose the trimer - (gT2)$_3$ as a reference molecule to compare with experimental results (the chemical structure of (gT2)$_3$ is shown in Figure S27). Figure S28 presents calculated spectra of (gT2)$_3$ in its neutral ('0'), singly ('+1') and doubly ('+2') oxidized states with Gaussian broadening equal 0.15 eV. Bottom panel of Figure S28 presents different linear combinations of these three spectra to illustrate some of the expected intermediate states. Table S1 presents calculated active excited states energies and corresponding oscillator strengths of (gT2)$_3$, (gT2)$_3^{+1}$ and (gT2)$_3^{+2}$.

To reduce computational time, n-type polymers p(g7NDI-gT2) and p(ZI-NDI-gT2) were approximated by their monomers with shorter side chains. Figure S29 shows structural formulas of monomers named gT-g1DI-gT and gT-ZI-NDI-gT, both consist of NDI core with ethylene glycol (g1NDI) or zwitterion (ZI-NDI) side chains and methoxy thiophene ring on each side of NDI (gT). Figure S30 and S31 show spectra of monomers in their neutral, singly and doubly reduced states with Gaussian broadening equal 0.25 eV. Bottom panels include spectra of partially charged species. Table S2 and S3 consist of list of excited states and corresponding oscillator strengths for gT-g1NDI-gT and gT-ZI-NDI-gT respectively.

Together with optical spectra, charge distributions across molecules in different oxidation states were calculated using CHEL PG scheme (Charges from Electrostatic Potentials using a Grid-based method) implemented in Gaussian16. In order to simplify the analysis of the results, molecules are divided into fragments, indicated with different colors in Figure S32 and Figure S33. Table S4 and Table S5 show total charge distributions for these fragments in neutral, polaron and bipolaron states of gT-g1NDI-gT and gT-ZI-NDI-gT respectively. Differences between neutral and polaron states and bipolaron and polaron states are also included in the tables.

As we comment in the main text of the paper, most of the added charge is distributed within the NDI core. In the case of gT-g1NDI-gT, 68% of the first extra electron (polaron) and 67% of the second extra electron (bipolaron) are located on the NDI. In the zwitterion case – gT-ZI-NDI-gT, the addition of the first extra electron causes an even stronger polarization in the side chains than in the neutral molecule and thus stronger charge localization on NDI – 85%. Including a second electron already breaks this trend. Only 58% of its charge is localized on NDI, the rest is spread among side chains and thiophenes.

In both molecules, the oxygen atoms in the NDI group are negatively charged in the neutral state and become even more negative upon reduction, much more than their neighboring positively charged carbons. This fact suggests that oxygens are favored locations for extra electrons in the first and second reductions of gT-g1NDI-gT and gT-ZI-NDI-gT and justifies the reduction mechanism proposed in reference [4].

## 9.1 p(gT2)

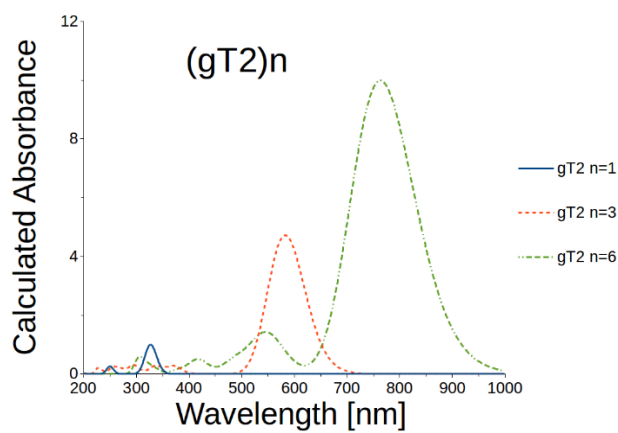

Figure S26. TD-DFT calculated absorption spectra of the monomer, trimer and hexamer (gT2)$_n$ in their neutral state normalized to the spectra of monomer. In order to reproduce experimental data on p(gT2), oligomers (gT2)$_n$ with n=1, 3 and 6 were calculated using above described procedure. For neutral systems, the lowest energy absorption peak position red-shifts with increasing number of monomer units from 328 nm for monomer, 583 nm for trimer to 763 nm for hexamer. The experimental value is around 645 nm as shown in Table 1 of the main text. On this basis, we chose the trimer - (gT2)$_3$ as a reference molecule to compare with experimental results (the chemical structure of (gT2)$_3$ is shown in Figure S27).

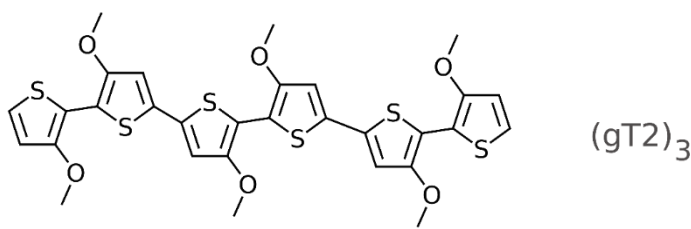

Figure S27. Chemical structure of the trimer (gT2)$_3$.

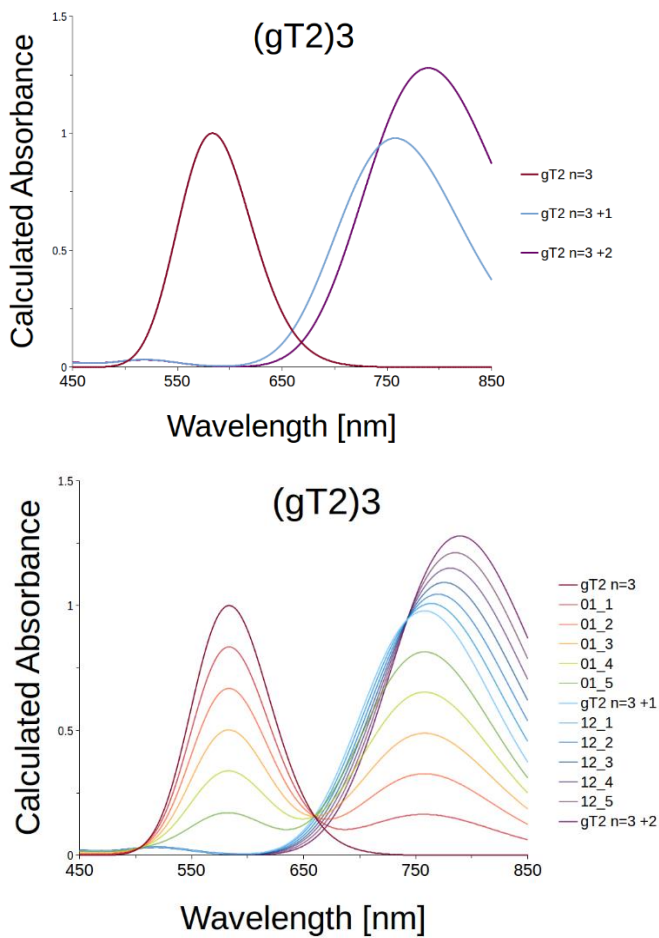

Figure S28. TD-DFT simulated absorption spectra of unchanged and charged state for the p(gT2) polymer. (top) TD-DFT calculated absorption spectra of the trimer (gT2)$_3$ in its neutral, singly and doubly oxidized state normalized to the spectra of neutral trimer. (bottom) Linear combination of the spectra qualitatively reproducing results shown in the main text for p(gT2). In the legend eg. '01_2' means the second (2) out of five step between neutral (0) and singly oxidized (1) trimer. Other names were created analogously.

Table S2. Energies and oscillator strengths calculated for (gT2)$_3$, (gT2)$_3^{+1}$ and (gT2)$_3^{+2}$.

| Neutral (gT2)$_3$ | | (gT2)$_3^{+1}$ | | (gT2)$_3^{+2}$ | |
|---|---|---|---|---|---|
| Energy [eV (nm)] | Oscillator strength | Energy [eV (nm)] | Oscillator strength | Energy [eV (nm)] | Oscillator strength |
| 2.12 (583.6) | 2.05 | 0.89 (1383.3) | 0.27 | 1.57 (789.4) | 2.62 |
| 3.19 (388.6) | 0.004 | 1.64 (758.1) | 2.01 | 2.37 (522.5) | 0.06 |
| 3.34 (370.9) | 0.12 | 2.20 (562.7) | 0.003 | 2.57 (481.6) | 0.02 |
| 3.67 (337.2) | 0.12 | 2.37 (522.5) | 0.02 | 2.77 (447.3) | 0.02 |
| | | 2.39 (517.8) | 0.05 | 2.87 (432.2) | 0.01 |
| | | 2.64 (470.4) | 0.01 | 3.01 (412.3) | 0.04 |
| | | 2.76 (449.8) | 0.03 | | |
| | | 3.02 (410.4) | 0.05 | | |

## 9.2 n-type polymer - p(g7NDI-gT2) and p(ZI-NDI-gT2)

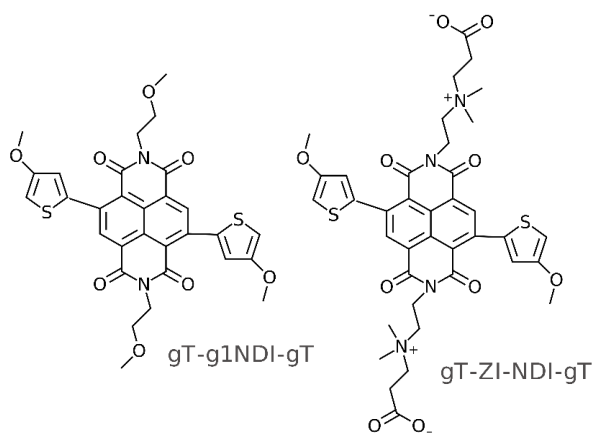

Figure S29. Structural formulas of calculated monomers of gT-g1NDI-gT and gT-ZI-NDI-gT.

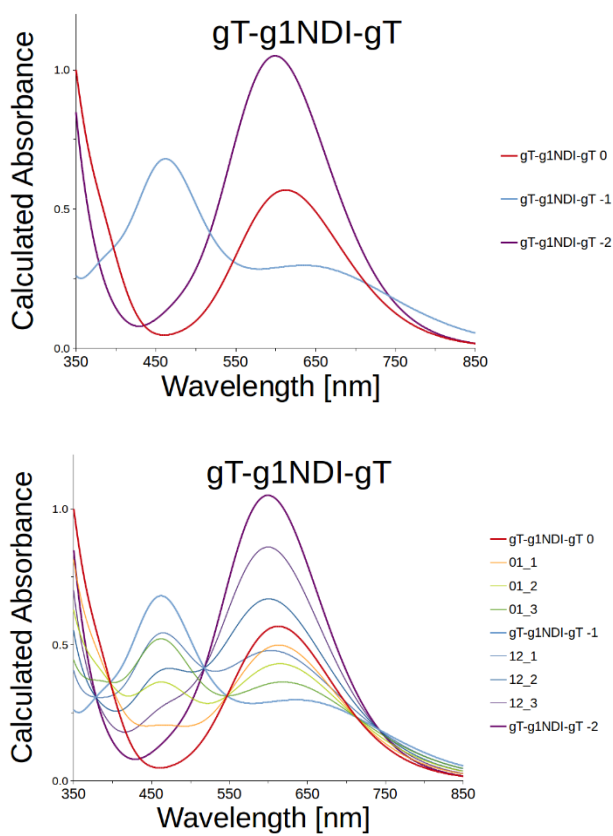

Figure S30. TD-DFT simulated absorption spectra of unchanged and charged state for the p(g7NDI-gT2) polymer. (top) TD-DFT calculated absorption spectra of the monomer gT-g1NDI-gT in its neutral, singly and doubly reduced state normalized to the spectra of neutral molecule. (bottom) Linear combination of the spectra qualitatively reproducing results shown in Figure S38 for p(g7NDI-gT2). In the legend eg. '01_2' means the second (2) out of three step between neutral (0) and singly reduced (1) molecules. Other names were created analogously.

Table S3. Energies and oscillator strengths calculated for gT-g1NDI-gT, gT-g1NDI-gT $^{-1}$ and gT-g1NDI-gT $^{-2}$.

| Neutral gT-g1NDI-gT | | gT-g1NDI-gT $^{-1}$ | | gT-g1NDI-gT $^{-2}$ | |
|---|---|---|---|---|---|
| Energy [eV (nm)] | Oscillator strength | Energy [eV (nm)] | Oscillator strength | Energy [eV (nm)] | Oscillator strength |
| 2.02 (613.7) | 0.25 | 1.80 (687.0) | 0.09 | 1.98 (627.2) | 0.01 |
| 2.21 (562.3) | 0.02 | 2.07 (598.2) | 0.06 | 2.07 (600.3) | 0.45 |
| 3.14 (394.7) | 0.03 | 2.36 (524.6) | 0.05 | 2.55 (486.7) | 0.06 |
| | | 2.58 (480.3) | 0.02 | 3.02 (410.6) | 0.01 |
| | | 2.68 (462.2) | 0.24 | | |
| | | 2.88 (430.4) | 0.02 | | |

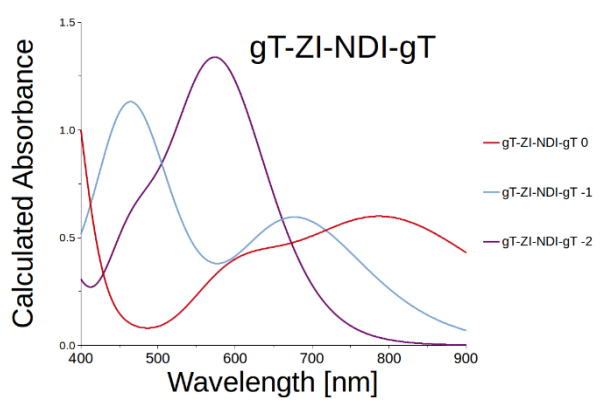

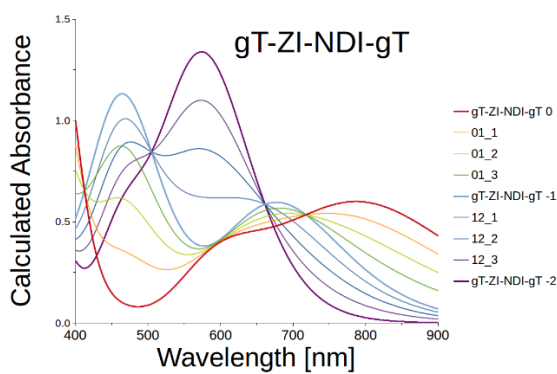

Figure S31. TD-DFT simulated absorption spectra of unchanged and charged state for the p(ZI-NDI-gT2) polymer (top) TD-DFT calculated absorption spectra of the monomer gT-ZI-NDI-gT in its neutral, singly and doubly reduced state normalized to the spectra of neutral molecule. (bottom) Linear combination of the spectra qualitatively reproducing results shown in the main text for p(ZI-NDI-gT2). In the legend eg. '01_2' means the second (2) out of three step between neutral (0) and singly reduced (1) molecules. Other names were created analogously.

Table S4. Energies and oscillator strengths calculated for gT-ZI-NDI-gT, gT-ZI-NDI-gT$^{-1}$ and gT-ZI-NDI-gT$^{-2}$.

| Neutral gT-ZI-NDI-gT | | gT-ZI-NDI-gT$^{-1}$ | | gT-ZI-NDI-gT$^{-2}$ | |
|---|---|---|---|---|---|
| Energy [eV (nm)] | Oscillator strength | Energy [eV (nm)] | Oscillator strength | Energy [eV (nm)] | Oscillator strength |
| 1.47 (843.4) | 0.05 | 1.70 (728.8) | 0.01 | 2.14 (579.2) | 0.35 |
| 1.57 (789.7) | 0.11 | 1.83 (679.2) | 0.16 | 2.21 (560.2) | 0.01 |
| 2.02 (613.4) | 0.09 | 2.38 (522.0) | 0.04 | 2.63 (471.9) | 0.16 |
| 2.07 (596.8) | 0.01 | 2.39 (518.8) | 0.01 | | |
| 2.27 (546.3) | 0.01 | 2.48 (500.5) | 0.03 | | |
| 2.69 (460.5) | 0.02 | 2.61 (474.9) | 0.03 | | |
| | | 2.66 (465.3) | 0.19 | | |
| | | 2.85 (435.0) | 0.05 | | |
| | | 2.93 (423.1) | 0.01 | | |
| | | 3.01 (411.4) | 0.05 | | |

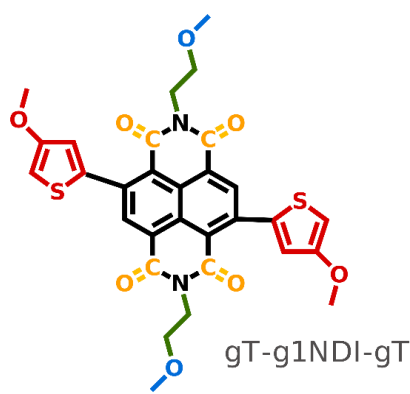

Figure S32. Molecules' sections of gT-g1NDI-gT proposed to clarify charge distribution presented in Table S4.

Table S5. Calculated charge distribution on selected fragments of gT-g1NDI-gT (colored in Figure S32). The values related to the neutral, polaron and bipolaron states are shown in columns 2, 3 and 5. Columns 4 and 6 consist of differential charge, being the difference in charge distribution between neutral and polaron states and bipolaron and polaron states respectively.

| Molecule's fragment | | gT-g1NDI-gT Neutral | gT-g1NDI-gT Polaron -1 | Charge difference between polaron and neutral state | gT-g1NDI-gT Bipolaron -2 | Charge difference between bipolaron and polaron state |
|---|---|---|---|---|---|---|
| $(CH_2O)_2$ | | -0.4 | -0.47 | -0.07 | -0.55 | -0.07 |
| $(C_2H_4)_2$ | | +0.73 | +0.74 | -0.01 | +0.73 | -0.00 |
| C=O | $C_4$ | +2.32 | +2.27 | -0.09 | +2.07 | -0.16 |
| | $O_4$ | -1.77 | -1.99 | -0.22 | -2.22 | -0.23 |

| | | | | | |
|---|---|---|---|---|---|
| NDI (without C=O) | -0.84 | -1.21 | -0.37 | -1.49 | -0.28 |
| (gT)₂ | -0.03 | -0.29 | -0.25 | -0.54 | -0.25 |

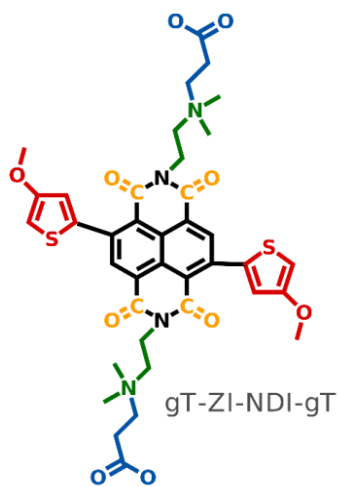

Figure S33. Molecules' sections of gT-ZI-NDI-gT proposed to clarify charge distribution presented in Table S5.

*Table S6. Calculated charge distribution on selected fragments of gT-ZI-NDI-gT (colored in Figure S33). The values related to the neutral, polaron and bipolaron states are shown in columns 2, 3 and 5. Columns 4 and 6 consist of differential charge that is the difference in charge distribution between neutral and polaron states and bipolaron and polaron states respectively.*

| Molecule's fragment | | gT-ZI-NDI-gT Neutral | gT-ZI-NDI-gT Polaron -1 | Charge difference between polaron and neutral state | gT-ZI-NDI-gT Bipolaron -2 | Charge difference between bipolaron and polaron state |
|---|---|---|---|---|---|---|
| $(C_3H_4O_2)_2$ | | -1.13 | -1.34 | -0.21 | -1.44 | -0.1 |
| $(C_4H_{10}N)_2$ | | +1.51 | +1.67 | +0.16 | +1.52 | -0.15 |
| C=O | $C_4$ | +2.52 | +2.45 | -0.18 | +2.08 | -0.37 |
| | $O_4$ | -1.89 | -2.1 | -0.21 | -2.27 | -0.17 |
| NDI (without C=O) | | -0.98 | -1.44 | -0.46 | -1.48 | -0.04 |
| (gT)₂ | | -0.02 | -0.23 | -0.21 | -0.41 | -0.17 |

# 10. Further electrochemical investigation and stability of p(gT2), p(ZI-NDI-gT2) and p(7NDI-gT2)

## 10.1 p(gT2)

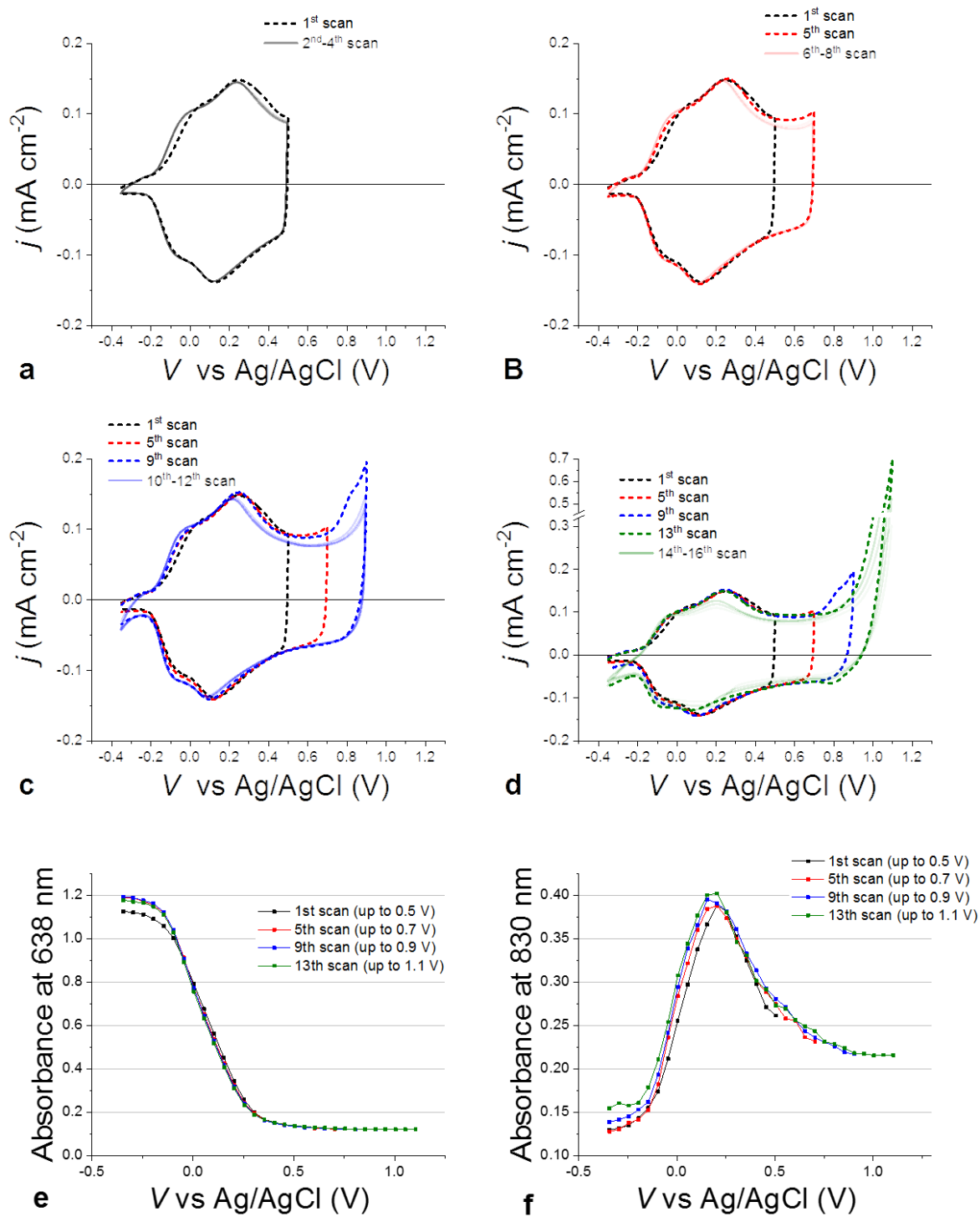

Figure S34. Electrochemical charging and reversibility of p(gT2) on FTO glass. (a) to (d) consecutive cyclic voltammetry measurements performed on a p(gT2) polymer film deposited on FTO in 0.1 M

NaCl:DIW at scan rate of 50 mV s$^{-1}$. Absorbance evolution monitored at (e) 638 nm and (f) 830 nm as a function of applied voltage during the CV scans shown in (a)-(d). The formation of bipolarons in the film, which can be monitored as the drop in absorbance at 830 nm after the first peak in (f), appears to complete at voltages of about 1.1 V.

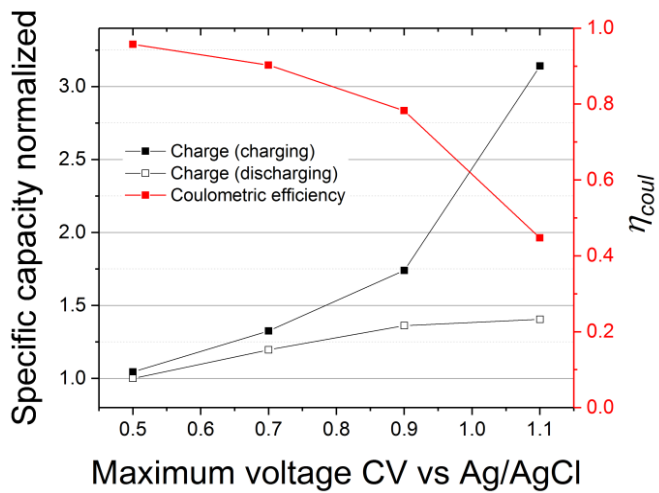

Figure S35. Specific capacity and coulombic efficiency of a p(gT2) film scanned up to different maximum potential vs Ag/AgCl of a cyclic voltammetry measurement.

## 10.2 p(g7NDI-gT2)

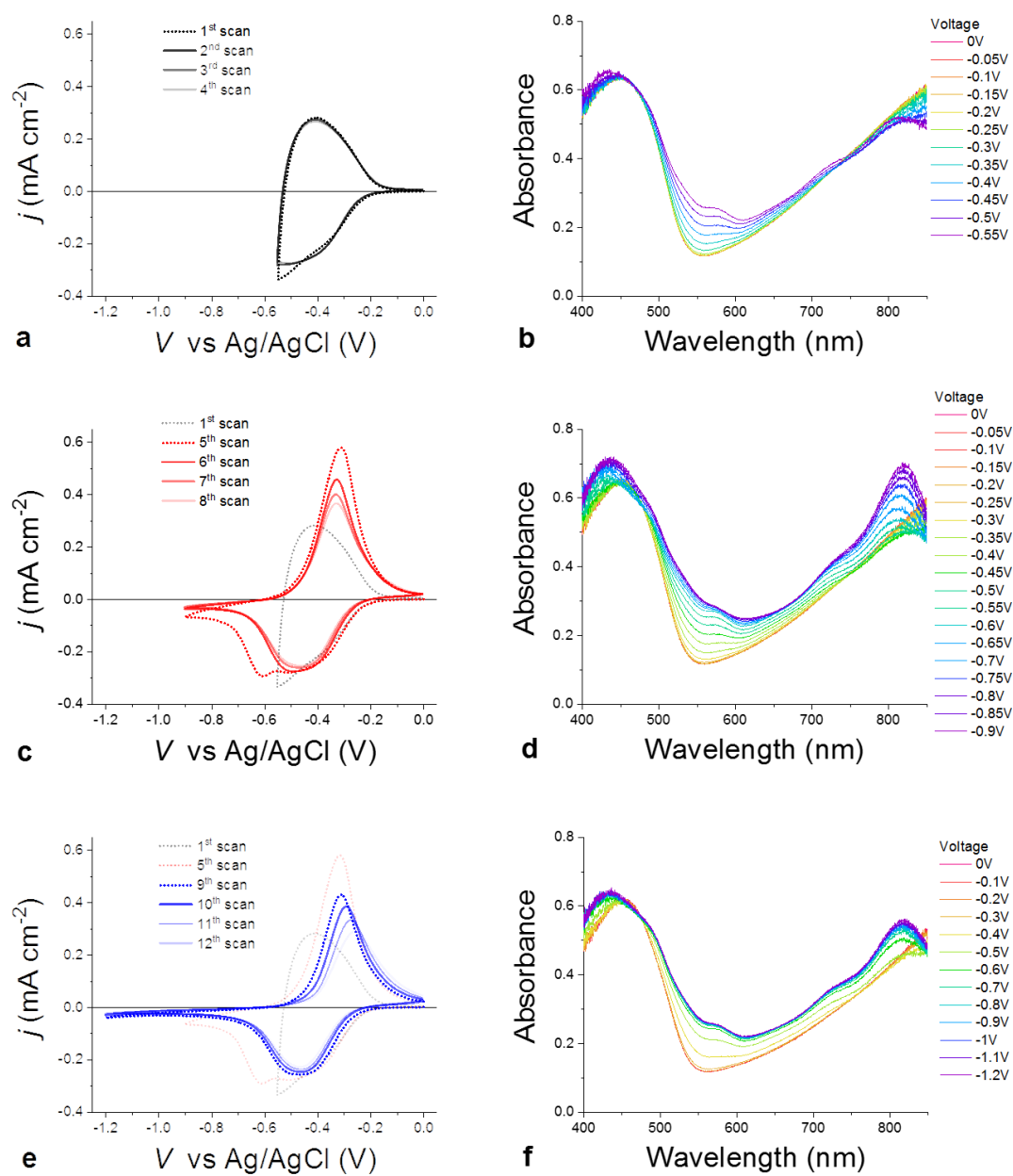

Figure S36. Reversibility and electrochemical charging of p(g7NDI-gT2) n-type polymer. (a), (c), (e) consecutive cyclic voltammetry measurements performed on a p(g7NDI-gT2) polymer film deposited on FTO in 0.1 M NaCl at scan rate of 50 mV s$^{-1}$ reaching a negative potential of (a) -0.55 V, (c) -0.9 V, (e) -1.2 V vs Ag/AgCl. Film absorbance evolution referred to the (b) 1$^{st}$, (d) 5$^{th}$ and (f) 9$^{th}$ scan. The presence of a second peak when scanning to potential more negative than -0.55 V vs Ag/AgCl is shown in (c). This peak is not there in the following scans as discussed in the main text. Also, the changes in absorbance detected for the first measurement scanning down to -0.9 V vs Ag/AgCl (5$^{th}$ scan in (c)) are greater than the ones appearing over the following cycles. This suggests that the ability of the film to reversibly reach deep level of charging changes after the first scan to potentials more negative than -0.55 V vs Ag/AgCl. The spectral signatures that we observe at potentials more negative than about -0.4 V vs Ag/AgCl could be ascribed to bipolaron bands.

## 10.3 Stability of p(gT2), p(ZI-NDI-gT2) and p(g7NDI-gT2)

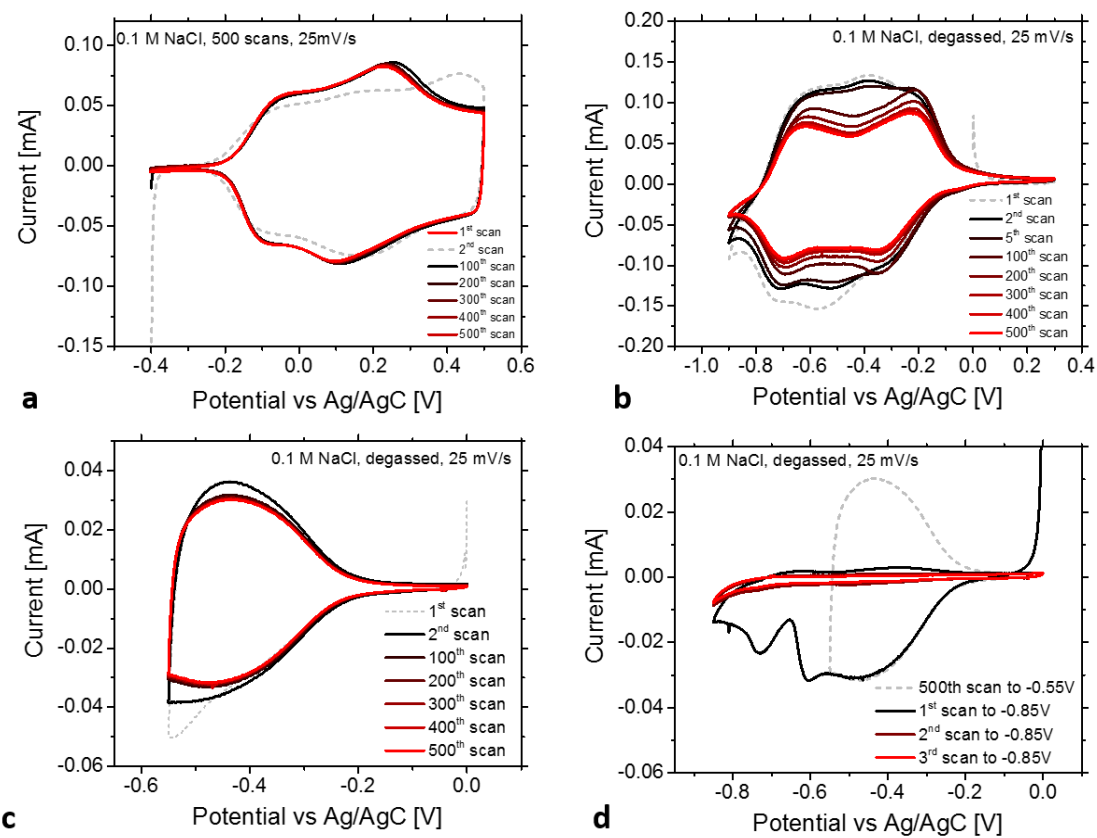

Figure S37. a) Stability measurements (500 scans) of p(gT2), b) p(ZI-NDI-gT2) , c) p(g7NDI-gT2) (-0.55V) and d) p(g7NDI-gT2) (-0.85V) on gold coated glass subtrates in 0.1 M NaCl with a scan rate of 25 mV s$^{-1}$ with a Ag/AgCl as the reference electrode. Degassed electrolytes were used for the n-type. The polymer p(g7NDI-gT2) is delimainting when charging to -0.85 V (the polymer film is the same as the one used to measure stability to -0.55 V).

## 10.4 Gravimetric capacity

The gravimetric capacities of p(gT2), p(ZI-NDI-gT2) and p(g7NDI-gT2) were measured by preparing solutions with known concentrations and the indicated volume of this solutions was drop cast on a gold substarte with a thin parylene interlayer to improve gold adhesion. For p(ZI-NDI-gT2), the substrate was heated to 80 °C during processing of the polymer, followed by a drying step at 140 °C for 30 mins. Note: The solubility of p(ZI-NDI-gT2) was limited at the chosen concentration and the gravimetric capacity is most likley underestimated.

p(gT2): 10 µL of a solution (2.2 mg/mL) was drop cast on Au coated glass substartes.

p(ZI-NDI-gT2): 10 µL of a solution (3.75 mg/mL) was drop cast on Au coated glass substartes.

p(g7NDI-gT2): 10 µL of a solution (2.00 mg/mL) was drop cast on Au coated glass substartes.

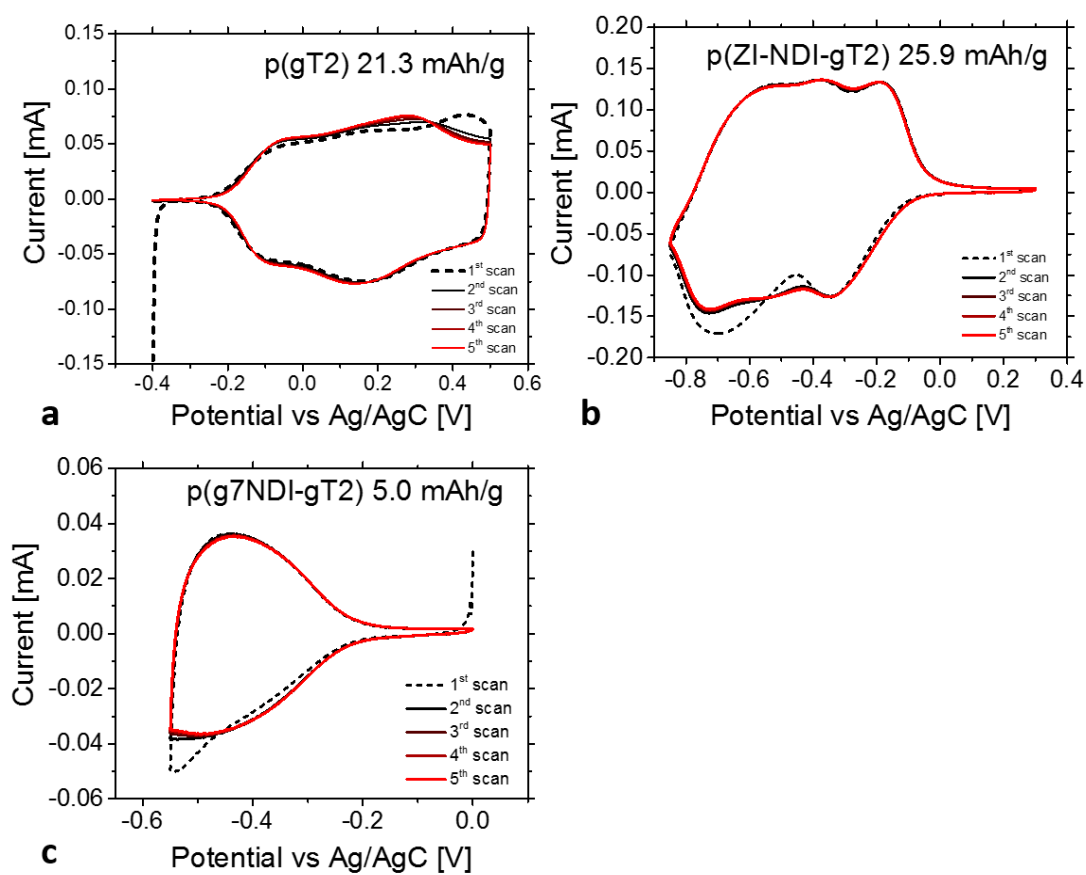

Figure S38. CV measurements of the prepared drop cast devices a) p(gT2) and b) p(ZI-NDI-gT2) and c) p(g7NDI-gT2) in degassed 0.1 M NaCl solution with a scan rate of 25 mV/s.

# 11. Charge density as a function of film thickness for the p-type and n-type polymers.

The thickness dependent charge density of the three polymers were extracted from cyclic voltammetry performed at 50 mV s$^{-1}$. The dotted lines are linear fits with slope 1 on a log-log scale to the data. For the p(gT2) and the p(g7NDI-gT2) polymers, the reversible charge of the second CV was taken for their respective datasets. For p(ZI-NDI-gT2), films deposited from methanol were used for this analysis. The reversible charge from the 1$^{st}$ CV was considered, probably representing an underestimate of the total charge that can be accumulated in films deposited from DMSO (the latter showed higher reversibility). The three datasets show a close to linear relation between charge density and thickness in the range 10 to 200 nm (the fits are run with slope constraint to 1), suggesting that charging and discharging occurs for all the polymers in the second timescale. The resulting specific capacities are shown in the legend.

These values are conservative estimates of the specific capacity for the polymers for the following reasons: (i) For the p(gT2) polymer a value of 25 mAh cm$^{-3}$ was found considering scanning of the polymer up to 0.5 V vs Ag/AgCl. Based on Figure S34 we can conclude that this is an underestimate of the potential specific capacity of p(gT2) polymer films in that larger values are expected when scanning to more positive potentials (up to 35 mAh cm$^{-3}$ when scanning up to 1.1 V vs Ag/AgCl). One expects however the coulombic efficiency to drop significantly under these conditions. (ii) For p(g7NDI-gT2) we consider only the charge accumulated in films scanned to -0.55V vs Ag/AgCl. (iii) For p(ZI-NDI-gT2), films deposited from methanol showed lower reversibility during the first CV scan compared to those deposited from DMSO. The theoretical capacity for the p(gT2) and the p(ZI-NDI-gT2) polymers can be calculated from the volume density of their individual redox-sites and the number of charges that each site can accommodate. Given the extended conjugation of these materials, the exact size of their redox sites is unknown. However, our experimental and computational results suggest that each monomer of the p(ZI-NDI-gT2) can theoretically be charged with two electrons yielding a theoretical capacity of about 51 mAh cm$^{-3}$. The value of 36 mAh cm$^{-3}$ that we observe is then about 70% of the material's theoretical capacity. For p(gT2), the value of 25 mAh cm$^{-3}$ that we measure corresponds to 0.6 holes per monomer unit. We expect that the polymer's individual redox sites involve more than one monomer. For example, if the polaron and bipolaron were to be delocalized on 2 or 3 monomer units, our result would suggest that a theoretical capacity of about 42 mAh cm$^{-3}$ or 28 mAh cm$^{-3}$ can be reached with this polymer. Note that these calculations have assumed a density of 1 g cm$^{-3}$ for the polymers.

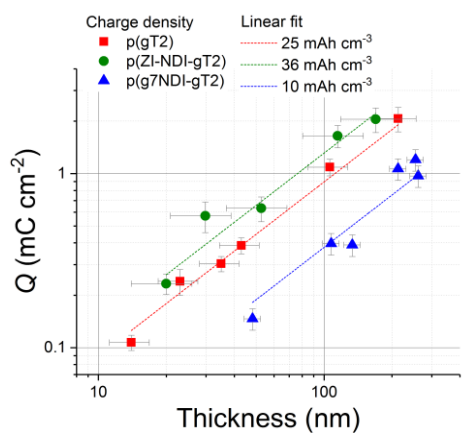

Figure S39. Charge density as a function of film thickness for the p-type and n-type polymers.

# 12. Rate-capabilities of the p-type and n type polymers

## 12.1 p(gT2)

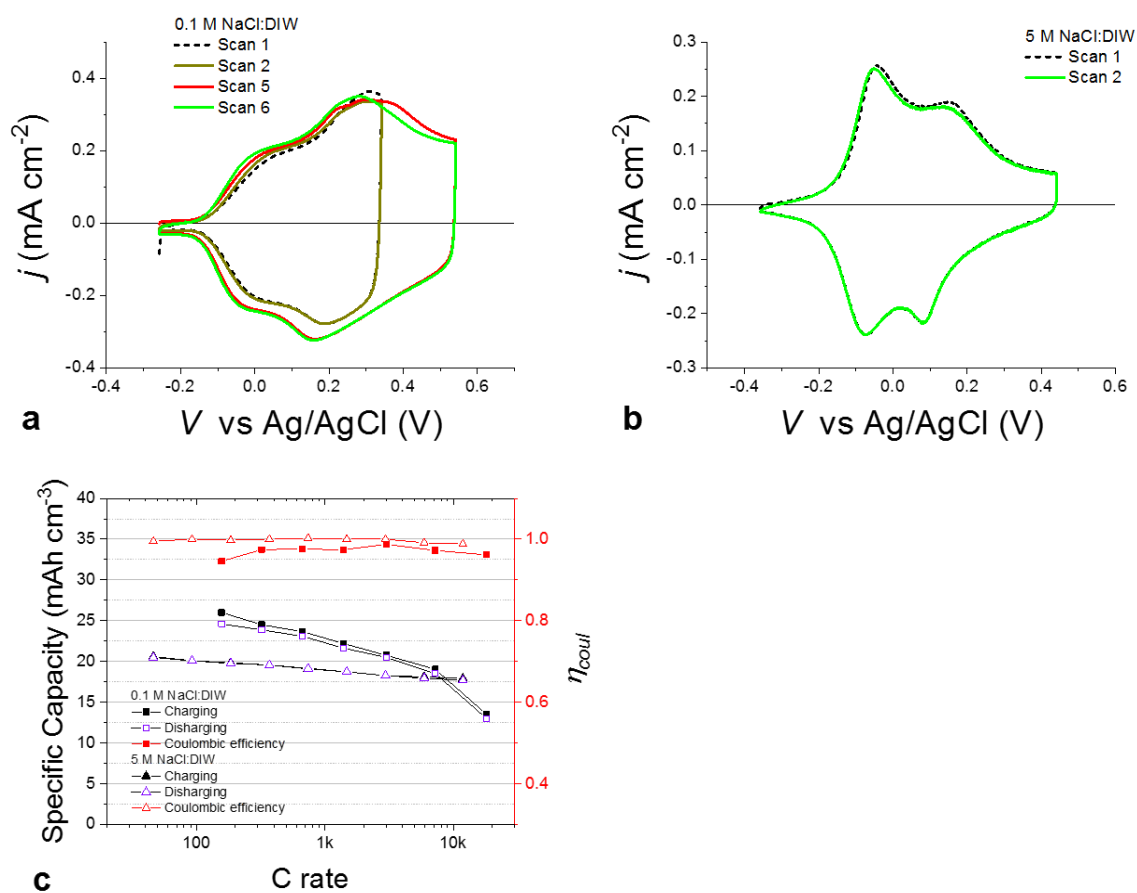

Figure S40. Cyclic voltammetry and Galvanostatic characterization for a p(gT2) polymer film. a)-b) Cyclic voltammetry performed at 50 mV s$^{-1}$ and c) Galvanostatic charging/discharging characterization of a 412 nm p(gT2) film deposited on gold coated glass. Characterization in aqueous 0.1 M NaCl was followed by rinsing of the film in DIW and characterization in aqueous 5 M NaCl. The fractional error associated to the value of specific capacity due to uncertainty in the measurement of the film volume is 22%.

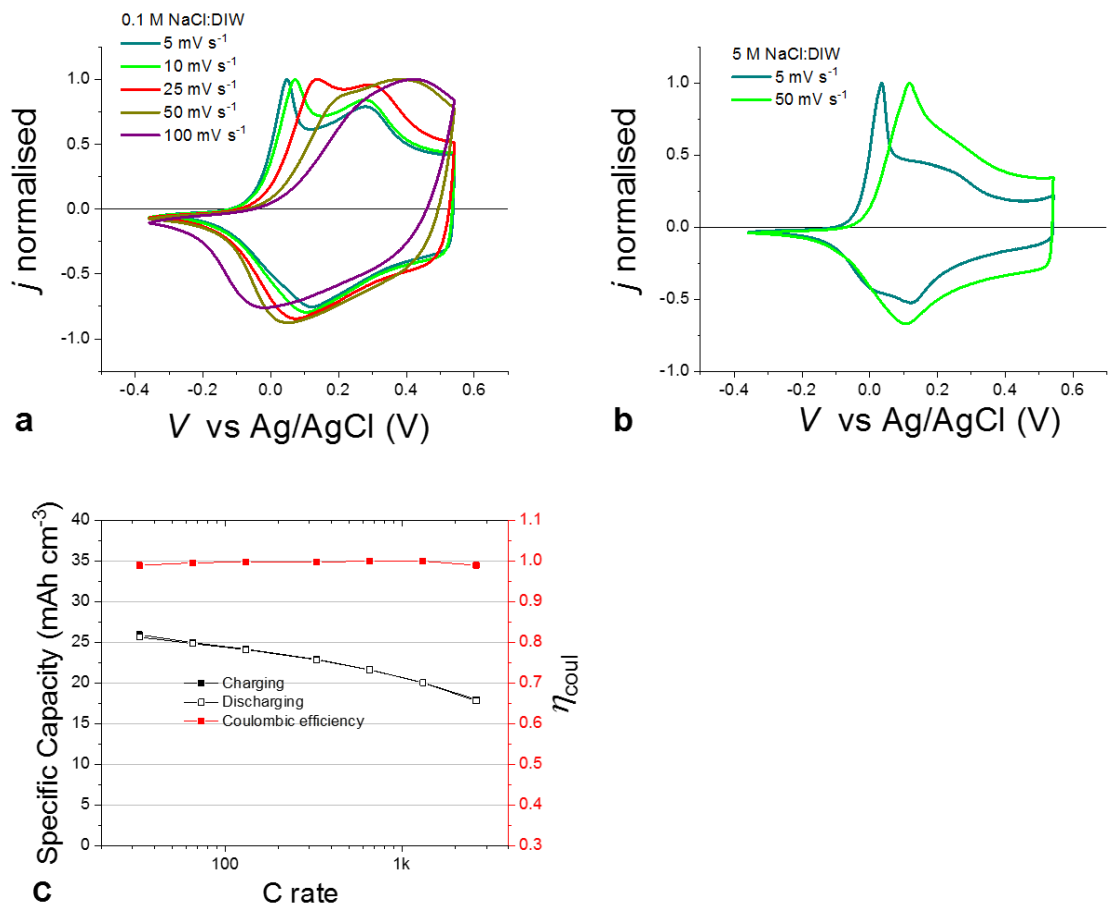

Figure S41. Cyclic voltammetry and Galvanostatic characterization for a p(gT2) polymer thick film. a)-b) Cyclic voltammetry and c) Galvanostatic charging/discharging characterization of a 8 μm p(gT2) film deposited on gold coated glass. Characterization in aqueous 0.1 M NaCl was followed by rinsing of the film in DIW and characterization in aqueous 5 M NaCl. Scan rate dependence of cyclic voltammetry in a) 0.1 M NaCl and b) 5 M NaCl DIW electrolytes. c) shows the rate capabilities of the film in 5 M NaCl aqeuous solution. The fractional error associated to the value of specific capacity due to uncertainty in the measurement of the film volume is 16%.

## 12.2 p(ZI-NDI-gT2)

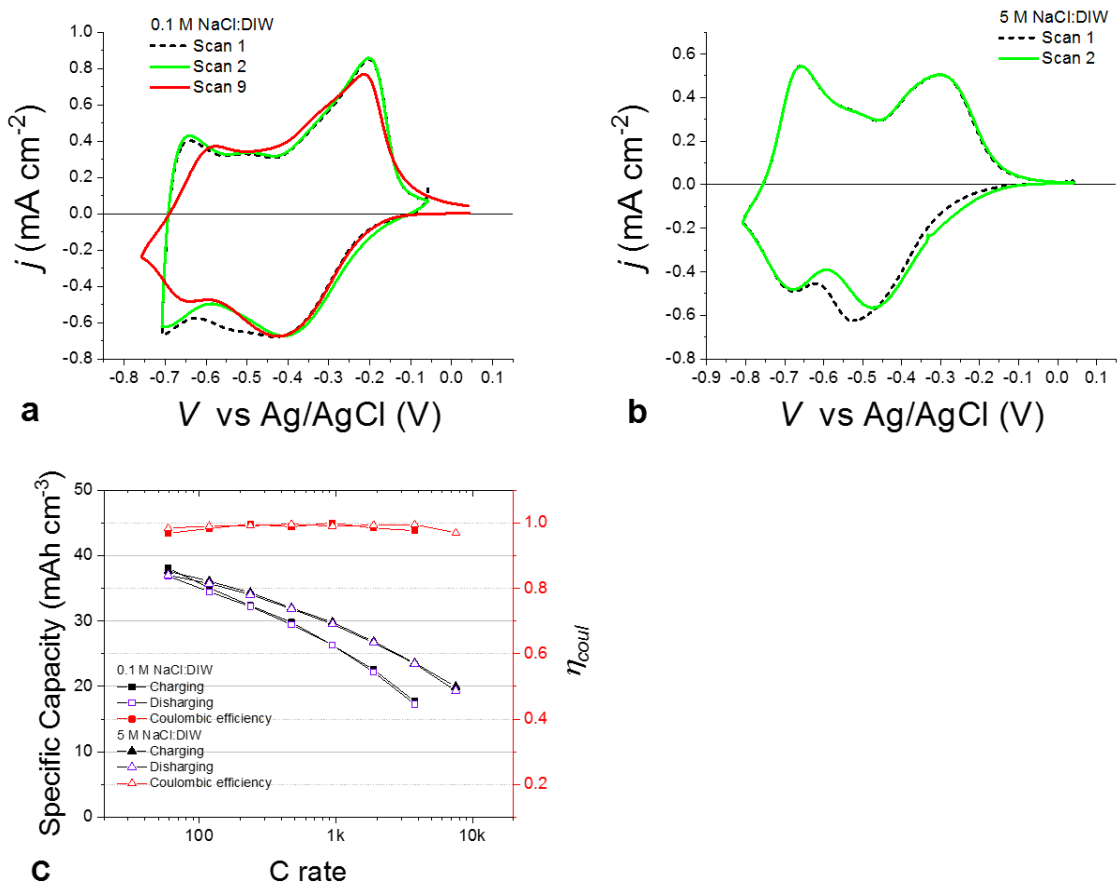

Figure S42. Cyclic voltammetry and Galvanostatic characterization for a p(ZI-NDI-gT2) polymer film. a)-b) Cyclic voltammetry performed at 50 mV s$^{-1}$ and c) Galvanostatic charging/discharging characterization of a 422 nm p(ZI-NDI-gT2) film deposited on gold coated glass. Characterization in aqueous 0.1 M NaCl was followed by rinsing of the film in DIW and characterization in aqueous 5 M NaCl. The fractional error associated to the value of specific capacity due to uncertainty in the measurement of the film volume is 20%.

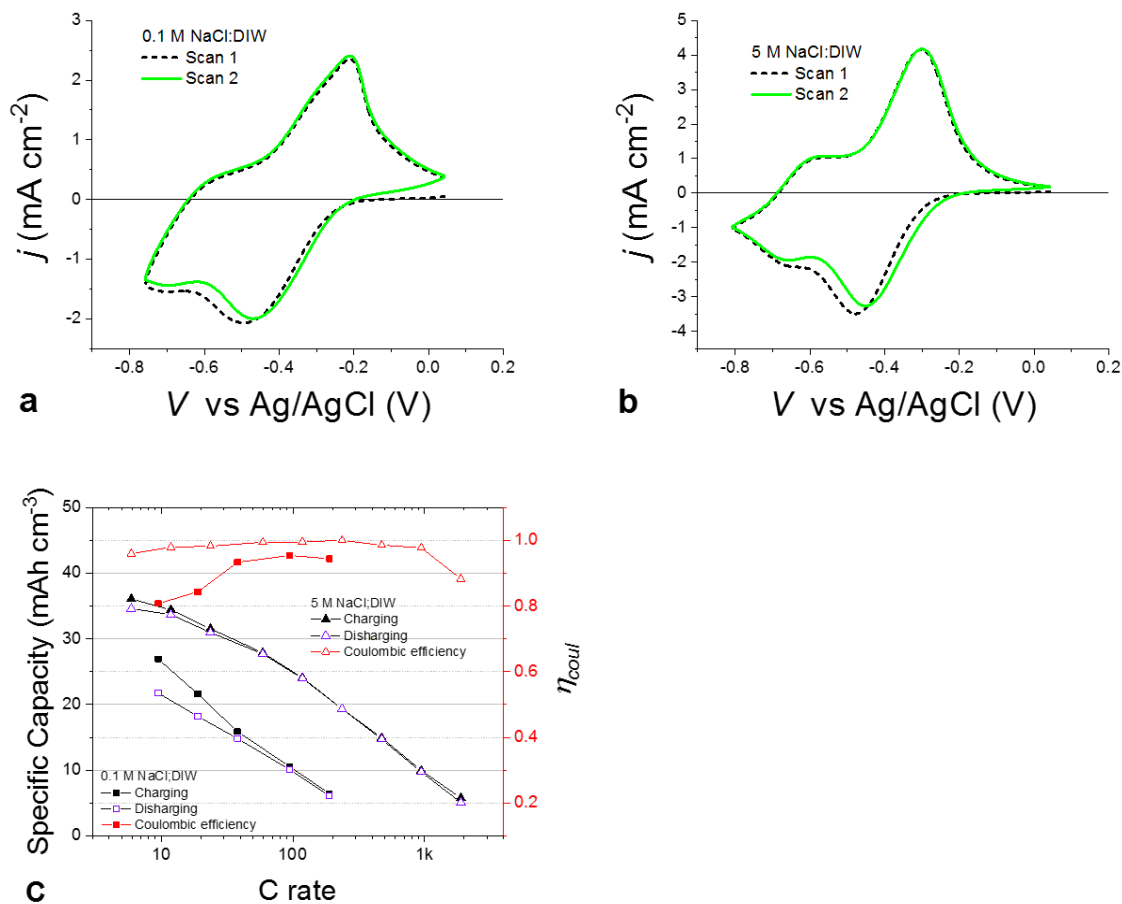

Figure S43. Cyclic voltammetry and Galvanostatic characterization for a p(ZI-NDI-gT2) polymer thick film. a)-b) Cyclic voltammetry performed at 50 mV s$^{-1}$ and (c) Galvanostatic charging/discharging characterization of a 2.6 µm p(ZI-NDI-gT2) film deposited on gold coated glass. Characterization in aqueous 0.1 M NaCl was followed by rinsing of the film in DIW and characterization in aqueous 5 M NaCl. The fractional error associated to the value of specific capacity due to uncertainty in the measurement of the film volume is 21%.

## 12.3 p(g7-NDI-gT2)

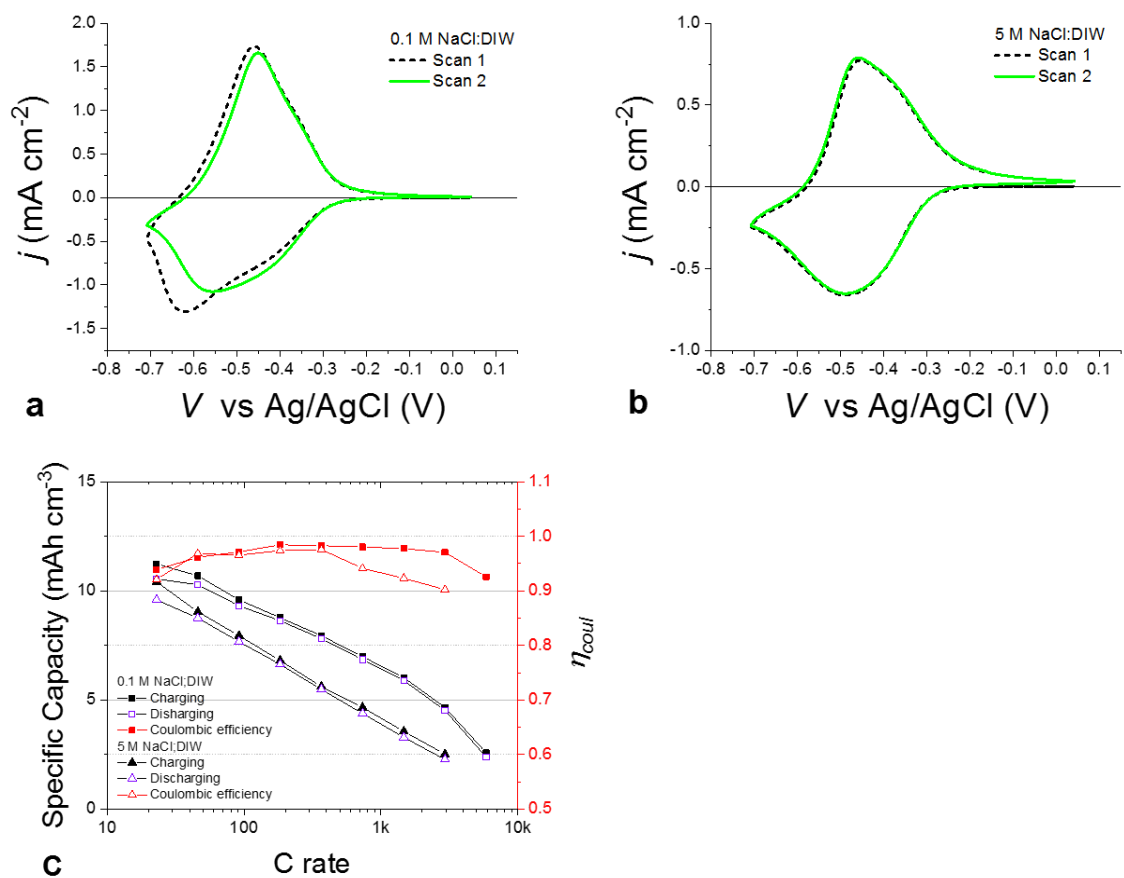

Figure S44. Thick film characterization for p(g7NDI-gT2) polymer. a)-b) Cyclic voltammetry performed at 50 mV s$^{-1}$ and c) Galvanostatic charging/discharging characterization of a 1.6 µm p(g7NDI-gT2) film deposited on gold coated glass. Characterization in aqueous 0.1 M NaCl was followed by rinsing of the film in DIW and characterization in aqueous 5 M NaCl. The fractional error associated to the value of specific capacity due to uncertainty in the measurement of the film volume is 21%.

# 13. Electrochemical cell

## 13.1 Voltage of the electrochemical cell

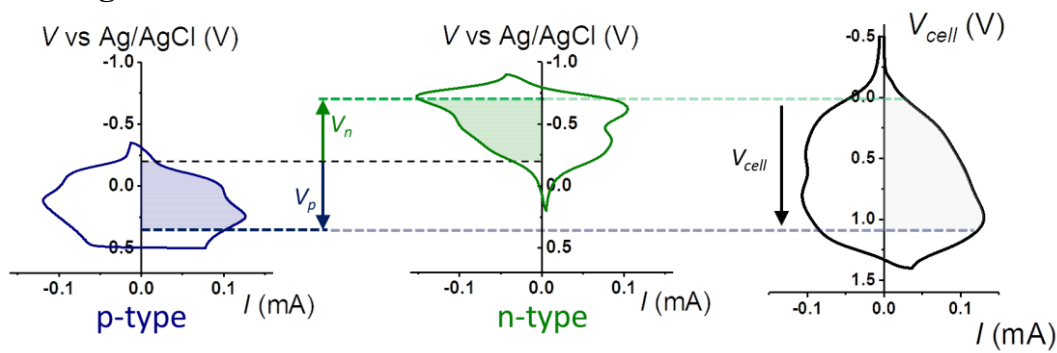

Figure S45. Energetics and charge distribution in the p- and n-type polymers measured in a two-electrode configuratrion: (left and center) cyclic voltammetry measurements on the p-type and the n-type polymers in a three electrode cell at 50 mV s$^{-1}$ (potentials are referenced to Ag/AgCl) and (right) cyclic voltammetry measurement on the complete battery at a scan rate of 100 mV s$^{-1}$. The value $V_{cell}$ is the difference in electrochemical potentials between the p-type (cathode) and n-type (anode) polymers.

## 13.2 Electrochemical response of the two terminal electrochemical cell using films with different thicknesses

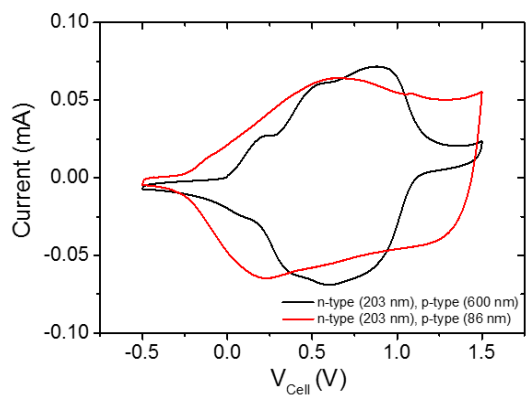

Figure S46. Influence of the p(gT2) film thickness on the electrochemical response of the two electrode cell measured with cyclic voltammetry.

## 13.3 Charge retention

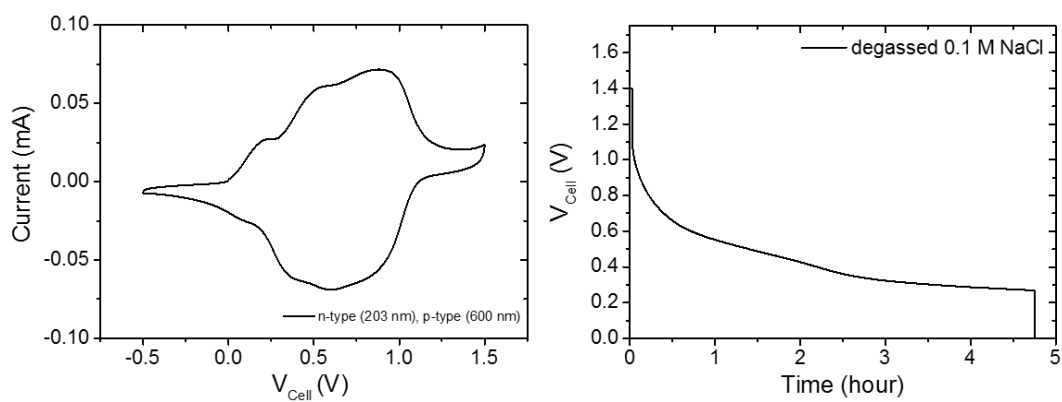

Figure S47. (a) CV of the electrochemical cell before the retention experiment. (b) Retention experiment where a voltage of 1.4 V is applied for 100 s to charge the cell followed by monitoring the potential for 286 min at open circuit. A degassed 0.1 M NaCl solution is used as the supporting electrolyte.

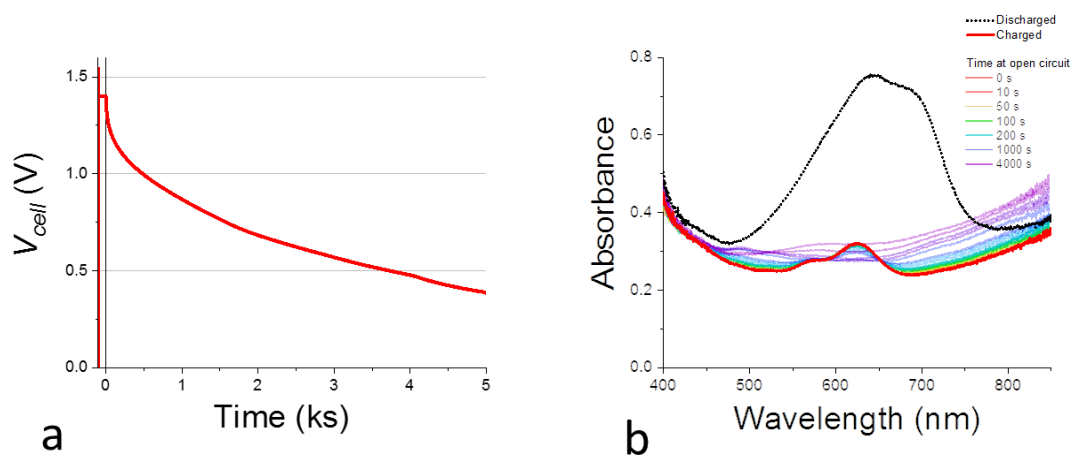

Figure S48. Charge retention experiment performed on the (p(gT2) / 0.1 M NaCl:DIW / p(ZI-NDI-gT2)) two terminal cell shown in the main text. (a) Voltage of the cell as a function of time: the voltage of the device was first kept at 1.4 V for 100 s (Time < 0 s); at Time = 0 s the device was switched to open circuit and its voltage was measured. (b) Optical absorbance of the two polymer films in series as a function of time during the retention measurement.

## 13.4 Hydrogen and oxygen detection

Thin films of p(gT2) and p(ZI-NDI-gT2) on ITO coated glass substrates were prepared and immersed in a 0.1 M degassed NaCl aqeuous solution. $H_2$ and $O_2$ were simultaneously detected in the electrolyte using two Clark electrodes ($H_2$-NP Unisense microsensor for hydrogen, OX-NP Unisense microsensor for oxygen, $O_2$ calibration for 0.1 M NaCl in $H_2O$ at 23 °C).[5] The electrolyte solution was purged with argon prior to the measurement until both electrodes showed a stable voltage output. An argon flow was then maintained in the gas phase of the cell for the first 2.82 h of the measurement in order to prevent oxygen from leaking back into the cell. The oxygen sensor was calibrated against argon-saturated and air-saturated electrolyte.

After stabilisation of the Clark electrode voltage output, the electrochemical cell was charged to 1.4 V for 31 min during which no hydrogen and oxygen evolution was detected (Figure S49a). In addition, no evidence of hydrogen and oxygen formation was observed when charging the cell for additonal 106 min after a 29 min period of no applied potential. After stopping the argon flow in the gas phase, the chronoamperometric current increased immediately, indicating that oxygen rapidy diffuses into the cell and reacts. Despite this current increase, a significant change of the oxygen concentration was only detected with a delay of 13 min after stopping the argon flow, which suggests that small initial amounts of oxygen are converted efficiently. As shown in section 15.5, hydrogen peroxide is a reaction product.

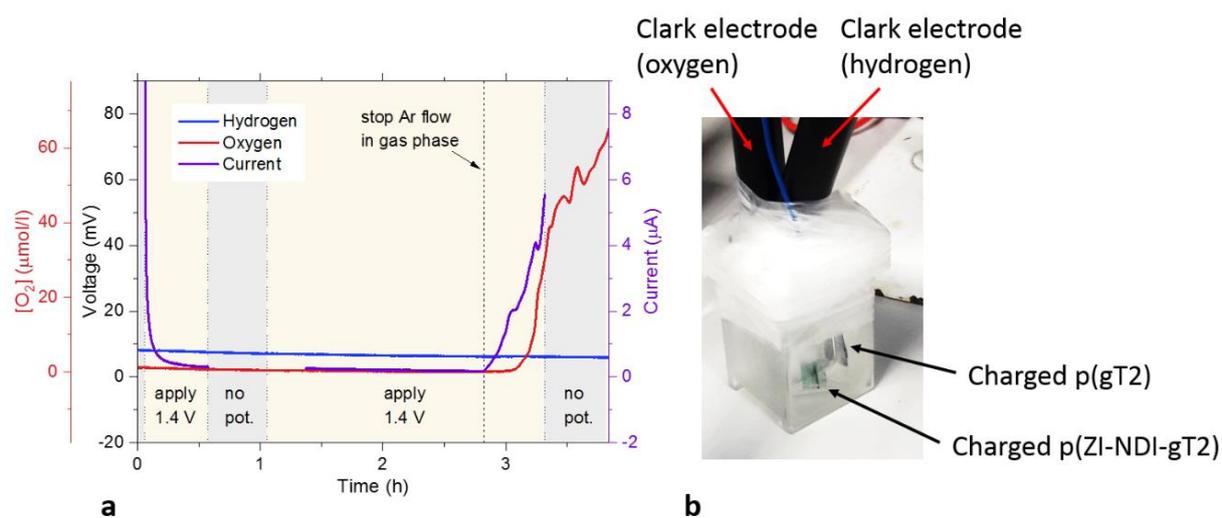

Figure S49. a) Detection of hydrogen and oxygen in 0.1 M NaCl and chronoamperometric current response under the different indicated conditions (1.4 V; no applied potential; with/without argon flow in the gas phase). b) Electrochemical cell setup including the hydrogen and oxygen sensitive Clark electrodes.

## 13.5 Hydrogen peroxide detection

Oxygen can be reduced to form hydrogen peroxide. There are two potential side reactions where oxygen can reduce p(gT2) or the reduced p(ZI-NDI-gT2) can reduce oxygen to become descharged.

Reduction of oxygen:

$$O_2 + 2H_3O^+ + 2e^- \rightarrow H_2O_2 + 2H_2O$$

Oxidation of p(gT2):

$$p(gT2) + H_3O^+ + \frac{1}{2}O_2 + NaCl \leftrightarrow p(gT2)^+\{Cl\}^- + \frac{1}{2}H_2O_2 + Na^+ + H_2O$$

Reduction of p(ZI-NDI-gT2):

$$p(ZINDIgT2)^-\{Na^+\} + H_3O^+ + \frac{1}{2}O_2 + Cl^- \leftrightarrow p(ZINDIgT2) + \frac{1}{2}H_2O_2 + NaCl + H_2O$$

Procedure for the detection of hydrogen peroxide with peroxidase/3,3',5,5'-tetramethylbenzidine

5.6 mg 3,3',5,5'-tetramethylbenzidine was dissolved in 10 mL of a 0.5 M HCl (dye solution). 5 mL of 0.1 M NaCl aqueous solution was used as the supporting electrolyte. After charging the electrochemical cell to 1.4 V for 200 seconds in ambient conditions, the solution was transfered into a cuvette (colorless solution, baseline). Then, 1 mL of a solution made of 0.04 mg/mL peroxidase from horseradish in deionised water was added and the spectra were recorded every 10 s.

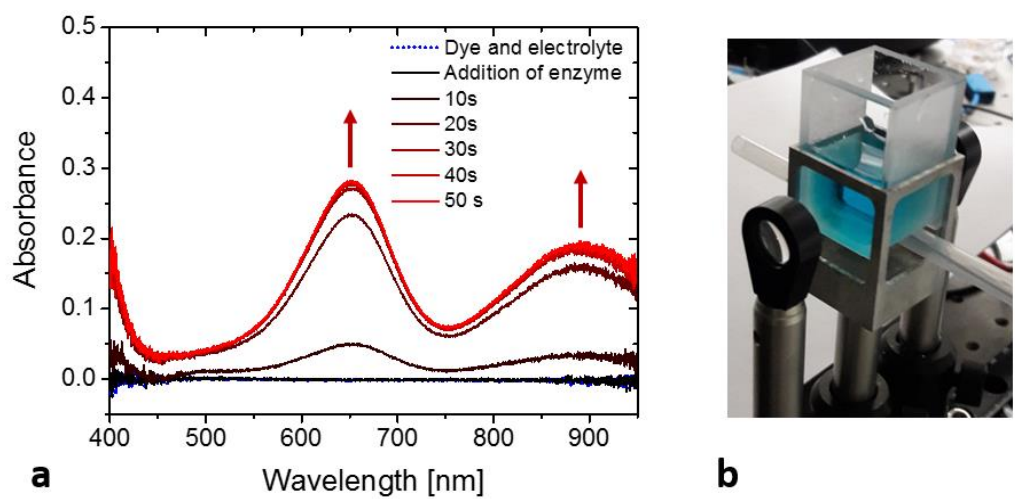

Figure S50. a) Color change of 3,3',5,5'-Tetramethylbenzidine in the presence of peroxidase and hydrogen peroxide formed during electrochemical charging of the cell in the presence of oxygen. b) color of the oxidised dye after addition of the enyzme.

## 13.6 Charging rate and stability of the electrochemical cell

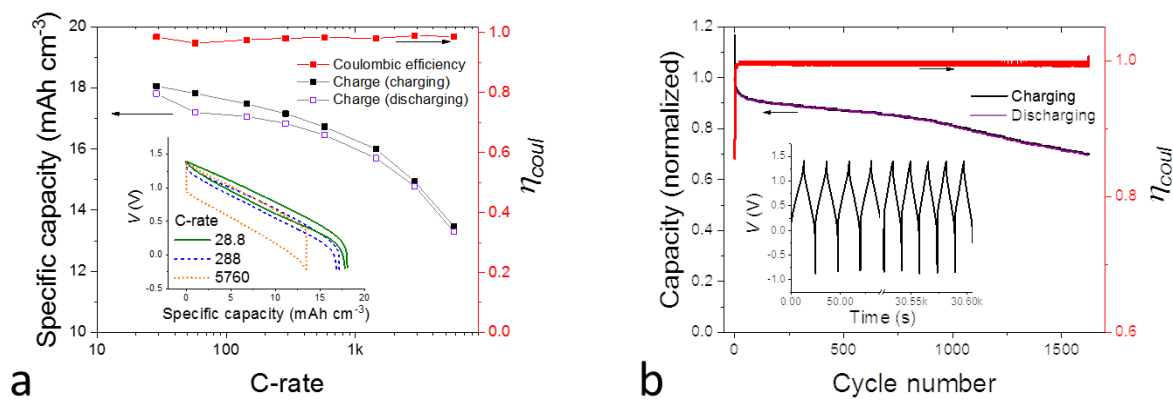

Figure S51. Galvanostatic and stability measurements on thin film p(gT2)/p(ZI-NDI-gT2) two electrode cell. (a) Specific capacity and coulombic efficiency as a function of C-rate (inset shows the galvanostatic charge-discharge curves at different C-rates). (b) Normalized specific capacity of the two-electrode cell and coulombic efficiency measured as a function of cycle number. The measurement was performed using Galvanostatic cycling at approximately 300 C-rate (inset showing the first and last cycles of the experiment).

## 13.7 Capacity recovery

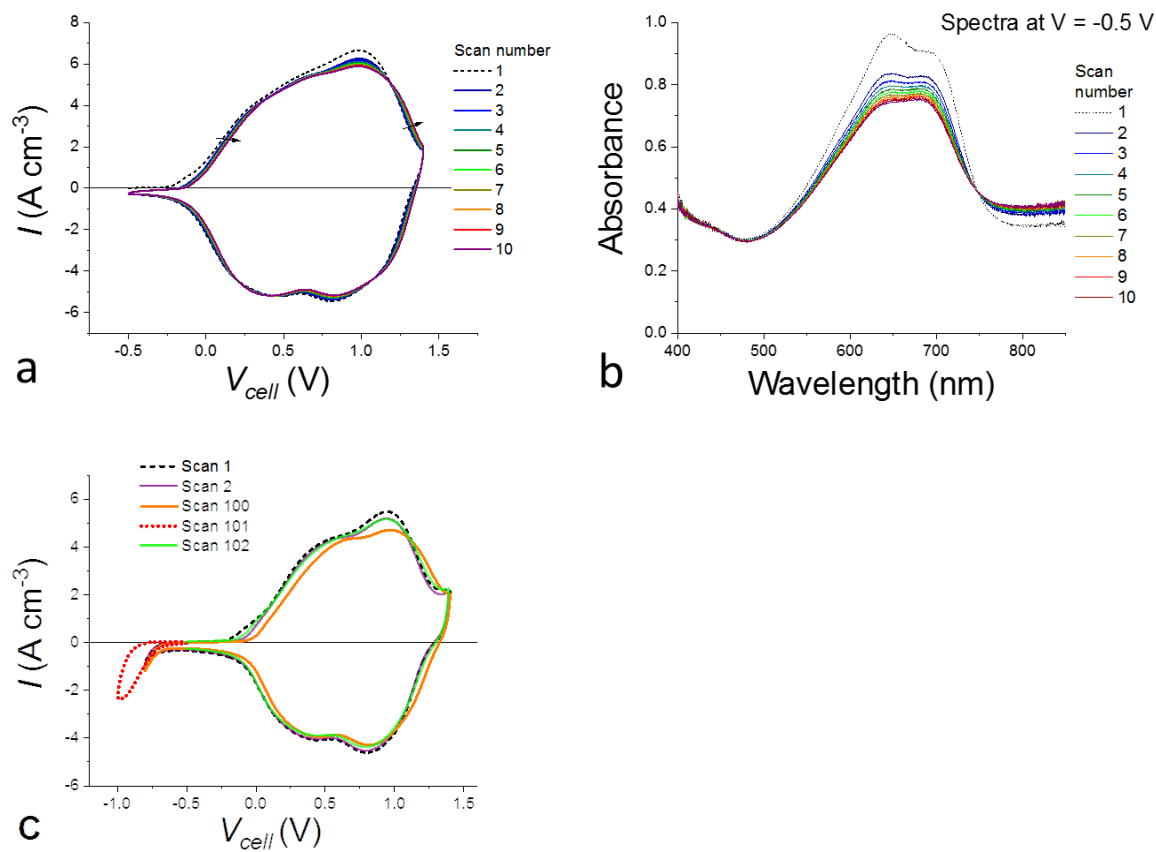

Figure S52. Probing retention limitations using spectroelectrochemistry. (a) Cyclic voltammetry measurements performed on the (p(gT2) / 0.1 M NaCl / p(ZI-NDI-gT2)) two electrode cell. (b) Optical absorbance of the two-electrode cell measured at the beginning of each scan. The features observed in cyclic voltammetry measurements shift in potential upon continuous cycling. The optical spectra shown in (b) suggests that the two polymers are in different charged stated at the beginning of each scan. We conclude that as a result of the loss of charges in the n-type polymer, the p-type polymer accumulates holes which cannot leave the film since there are not enough electrons on the anode of the cell. (c) Capacity recovery for the electrochemcal cell obtained after 100 cycles by applying a negative voltage at which p(gT2) can be reduced. The next measurement (scan 102) shows a cyclic voltammogram that is comparable to the second cycle.

## Supporting References